\documentclass[twocolumn]{svjour3}                          % two columns
\pdfoutput=1
\usepackage{amsmath}                                        % For align
\usepackage{preprintcover}                                  % To provide an ATLAS cover
\usepackage{booktabs}                                       % Nice tables
\usepackage{cite}                                           % For abbreviated citation lists
\usepackage{fix-cm}                                         % Fix Computer Modern font sizes
\usepackage{functions}                                      % Useful functions/abbreviations
\usepackage{graphicx}                                       % Add functionality to includegraphics
\usepackage{grffile}                                        % For special characters in filenames
\usepackage[hyperindex,breaklinks]{hyperref}                % For hyperlinks
\usepackage{mathptmx}                                       % use Times fonts if available
\usepackage[subrefformat=parens,labelformat=parens]{subfig} % For subfloat

\DeclareRobustCommand{\thetitle}{Measurements of jet vetoes and azimuthal decorrelations in \dijet events produced in $pp$ collisions at $\sqrt{s}=\unit{7}{\TeV}$ using the ATLAS detector}
\DeclareRobustCommand{\thejournal}{The European Physical Journal C}

\PreprintCoverPaperTitle{\thetitle}%
\PreprintIdNumber{CERN-PH-EP-2014-132}%
\PreprintJournalName{\thejournal}%

\PreprintCoverAbstract{Additional jet activity in \dijet events is measured using \maths{pp} collisions at ATLAS at a centre-of-mass energy of \unit{7}{\TeV}, for jets reconstructed using the \antikt algorithm with radius parameter \maths{R=0.6}.
This is done using variables such as the fraction of \dijet events without an additional jet in the rapidity interval bounded by the \dijet subsystem and correlations between the azimuthal angles of the \dijet{s}.
They are presented, both with and without a veto on additional jet activity in the rapidity interval, as a function of the mean transverse momentum of the \dijet{s} and of the rapidity interval size.
The double differential \dijet \xs is also measured as a function of the interval size and the azimuthal angle between the \dijet{s}.
These variables probe differences in the approach to resummation of large logarithms when performing QCD calculations.
The data are compared to \Powheg, interfaced to the \PythiaEight and \Herwig parton shower generators, as well as to \HEJ with and without interfacing it to the \Ariadne parton shower generator.
None of the theoretical predictions agree with the data across the full phase-space considered; however, \PowhegP and \HEJA are found to provide the best agreement with the data.
These measurements use the full data sample collected with the ATLAS detector in \unit{7}{\TeV} \maths{pp} collisions at the LHC and correspond to integrated luminosities of \unit{36.1}{\ipb} and \unit{4.5}{\ifb} for data collected during 2010 and 2011 respectively.

}%

\begin{document}\sloppy % relax rules for inter-word spacing

\title{\thetitle}
\author{The ATLAS Collaboration}
\institute{CERN, 1211 Geneva 23, Switzerland}
\journalname{\thejournal}
\renewcommand*{\today}{July 21, 2014}
\date{\today}

\maketitle

\begin{abstract}
Additional jet activity in \dijet events is measured using \maths{pp} collisions at ATLAS at a centre-of-mass energy of \unit{7}{\TeV}, for jets reconstructed using the \antikt algorithm with radius parameter \maths{R=0.6}.
This is done using variables such as the fraction of \dijet events without an additional jet in the rapidity interval bounded by the \dijet subsystem and correlations between the azimuthal angles of the \dijet{s}.
They are presented, both with and without a veto on additional jet activity in the rapidity interval, as a function of the mean transverse momentum of the \dijet{s} and of the rapidity interval size.
The double differential \dijet \xs is also measured as a function of the interval size and the azimuthal angle between the \dijet{s}.
These variables probe differences in the approach to resummation of large logarithms when performing QCD calculations.
The data are compared to \Powheg, interfaced to the \PythiaEight and \Herwig parton shower generators, as well as to \HEJ with and without interfacing it to the \Ariadne parton shower generator.
None of the theoretical predictions agree with the data across the full phase-space considered; however, \PowhegP and \HEJA are found to provide the best agreement with the data.
These measurements use the full data sample collected with the ATLAS detector in \unit{7}{\TeV} \maths{pp} collisions at the LHC and correspond to integrated luminosities of \unit{36.1}{\ipb} and \unit{4.5}{\ifb} for data collected during 2010 and 2011 respectively.

\end{abstract}

\section{Introduction}\label{sec:intro}
The Large Hadron Collider (LHC) has opened up a new kinematic regime to test perturbative QCD (pQCD) using measurements of jet production.
Next-to-leading-order QCD predictions for inclusive jet and \dijet \xs{s} have been found to describe the data at the highest measured energies~\cite{Aad:2010ad,Chatrchyan:2011qta,CMS:2011ab,Aad:2011fc,Chatrchyan:2012bja,Aad:2013lpa,Aad:2013tea}.
However, purely fixed-order calculations are expected to describe the data poorly wherever higher-order corrections to a given observable are important.
In such cases, higher orders in perturbation theory must be resummed; this resummation is typically performed in terms of \maths{\ln(1/x)}, where \maths{x} is Bjorken-$x$, the Balitsky--Fadin--Kuraev--Lipatov (BFKL) approach~\cite{Lipatov:1976zz,Kuraev:1976ge,Kuraev:1977fs,Balitsky:1978ic}, or in terms of \maths{\ln(Q^2)}, where \maths{Q^2} is the virtuality of the interaction, the Dokshitzer--Gribov--Lipatov--Altarelli--Parisi (DGLAP) approach~\cite{Gribov:1972ri,Altarelli:1977zs,Dokshitzer:1977sg}.
These resummations provide approximations that are most valid in phase-space regions for which the resummed terms provide a dominant contribution to the observable.
Such a situation exists in \dijet topologies when the two jets have a large rapidity separation or when a veto is applied to additional jet activity in the rapidity interval bounded by the \dijet system~\cite{DuranDelgado:2011tp}.
In these regions of phase-space higher order corrections proportional to the rapidity separation and the logarithm of the average transverse momentum of the dijets become increasing important: these must be summed to all orders to obtain accurate theoretical predictions.

When studying these phase-space regions, a particularly interesting observable is the gap fraction, \maths{f\left(\Qnought\right)}, defined as \maths{f\left(\Qnought\right) = \sigma_{\mathrm{jj}}\left(\Qnought\right)/\sigma_{\mathrm{jj}}} where \maths{\sigma_{\mathrm{jj}}} is the inclusive \dijet \xs and \maths{\sigma_{\mathrm{jj}}\left(\Qnought\right)} is the \xs for \dijet production in the absence of jets with transverse momentum greater than \Qnought in the rapidity interval bounded by the \dijet system.
The variable \Qnought is referred to as the veto scale.
In the limit of large rapidity separation, \DeltaY, between the jet centroids, the gap fraction is expected to be sensitive to BFKL dynamics~\cite{Andersen:2009nu,Andersen:2010ih,Forshaw:2005sx}.
Alternatively, when the average transverse momentum of the \dijet{s}, \pTbar, is much larger than the veto scale, the effects of wide-angle soft gluon radiation may become important~\cite{Forshaw:2006fk,Forshaw:2009fz,Forshaw:2012bi}.
Finally, \dijet production via \maths{t}-channel colour-singlet exchange~\cite{Mueller:1992pe} is expected to provide an increasingly important contribution to the total \dijet \xs when both of these limits are approached simultaneously.
The mean number of jets above the veto scale in the rapidity interval between the \dijet{s} is presented as an alternative measurement of hard jet emissions in the rapidity interval.

A complementary probe of higher-order QCD effects can be made by studying the azimuthal angle between the jets in the \dijet system, \DeltaPhi.
A purely \maths{2\rightarrow2} partonic scatter produces final-state partons back-to-back in azimuthal angle.
Any additional quark or gluon emission alters the balance between the partons and produces an azimuthal decorrelation, the predicted magnitude of which is different for fixed-order calculations, BFKL-inspired resummations and DGLAP-inspired resummations~\cite{Marquet:2007xx}.
In particular, the azimuthal decorrelation is expected to increase with increasing rapidity separation if BFKL effects are present~\cite{Vera:2007kn,Angioni:2011wj}.
To discriminate between DGLAP-like and BFKL-like behaviour, the azimuthal angular moments \maths{\langle \cos \left( n \left(\pi - \Delta \phi\right) \right) \rangle} where \maths{n} is an integer and the angled brackets indicate the profiled mean over all events, have been proposed~\cite{Vera:2006un,Vera:2007kn,Angioni:2011wj}.
In addition, taking the ratio of different angular moments is predicted to enhance BFKL effects~\cite{Ducloue:2012bm,Ducloue:2013hia}.

Previous measurements have been made of \dijet production for which a strict veto, of order of \maths{\Lambda_\text{QCD}}, was imposed on the emission of additional radiation in the \interjet region, so-called ``rapidity gap'' events, at HERA~\cite{Derrick:1995pb,Adloff:2002em,Chekanov:2006pw} and at the Tevatron~\cite{Abachi:1994hb,Abe:1994de,Abe:1997ie,Abe:1998ip,Abbott:1998jb}.
At the LHC, measurements of forward rapidity gaps and dijet production have been made by ATLAS~\cite{Aad:2011jz,Aad:2012pw}, while ratios of exclusive-to-inclusive \dijet \xs{s} have been measured at CMS~\cite{Chatrchyan:2012pb}.
Azimuthal decorrelations for central \dijet{s} have also been measured at the LHC by ATLAS~\cite{daCosta:2011ni} and CMS~\cite{Khachatryan:2011zj} and before that by D0 at the Tevatron~\cite{Abazov:2004hm}.

This paper presents measurements of the gap fraction and the mean number of jets in the rapidity interval as functions of both the \dijet rapidity separation and the average \dijet transverse momentum.
Measurements of the first azimuthal angular moment, the ratio of the first two moments and the double-differential \xs{s} as functions of \DeltaPhi and \DeltaY are also presented, both for an inclusive \dijet sample and for events where a jet veto is imposed.
Previous results are extended out to a \dijet rapidity separation of \maths{\DeltaY = 8} as well as to average \dijet transverse momentum of \maths{\pTbar = \unit{1.5}{\TeV}}, the effective kinematic limits of the ATLAS detector for \maths{pp} collisions at a centre-of-mass energy, \maths{\sqrt{s}=\unit{7}{\TeV}}.
The measurements are obtained using the full \maths{pp} collision datasets recorded during 2010 and 2011, corresponding to integrated luminosities of \unit{\maths{36.1\pm1.3}}{\ipb} and \unit{\maths{4.5\pm0.1}}{\ifb}, respectively~\cite{Aad:2013ucp}.
The two datasets are used in complementary areas of phase-space: the data collected during 2010 are used in the full rapidity range covered by the detector, probing large rapidity separations, with a veto scale of \unit{20}{\GeV}, while the data collected during 2011 are used in a restricted rapidity range and a veto scale of \unit{30}{\GeV} but can access higher values of \dijet transverse momentum.

The content of the paper is as follows.
\SectionRef{sec:detector} describes the ATLAS detector followed by \sectionRef{sec:mc}, which details the \MC simulation samples used.
Jet reconstruction and event selection are presented in \sectionRef{sec:jet-reconstruction} and \sectionRef{sec:event-selection} respectively.
The correction for detector effects is shown in \sectionRef{sec:unfolding} and discussion of systematic uncertainties on the measurement is in \sectionRef{sec:systematics}.
\SectionRef{sec:theory} discusses the theoretical predictions before the results are presented in \sectionRef{sec:results}.
Finally, the conclusions are given in \sectionRef{sec:summary}.

\section{The ATLAS detector}\label{sec:detector}
ATLAS~\cite{Aad:2008zzm} is a general-purpose detector surrounding one of the interaction points of the LHC.
The main detector components relevant to this analysis are the inner tracking detector and the calorimeters; in addition, the minimum bias trigger scintillators (MBTS) are used for selecting events during early data taking.
The inner tracking detector covers the pseudorapidity range \maths{\absEta<2.5}\footnote{ATLAS uses a right-handed coordinate system with its origin at the nominal interaction point (IP) in the centre of the detector and the \maths{z}-axis along the beam pipe. The \maths{x}-axis points from the IP to the centre of the LHC ring, and the \maths{y}-axis points upward. Cylindrical coordinates \maths{(r,\phi)} are used in the transverse plane, \maths{\phi} being the azimuthal angle around the beam pipe. The pseudorapidity is defined in terms of the polar angle \maths{\theta} as \maths{\eta=-\ln\tan(\theta/2)}. The rapidity of a particle with respect to the beam axis is defined as \maths{y = (1/2)\ln\left[(E+p_z)/(E-p_z)\right]}.} and has full coverage in azimuthal angle.
There are three main components to the inner tracker.
In order, moving outwards from the beam-pipe, these are the silicon pixel detector, the silicon microstrip detector and the straw-tube transition-radiation tracker.
These components are arranged in concentric layers and immersed in a \unit{2}{\tesla} magnetic field provided by the superconducting solenoid magnet.

The calorimeter is also divided into sub-detectors, providing overall coverage for \maths{\absEta<4.9}.
The electromagnetic calorimeter, covering the region \maths{\absEta<3.2}, is a high-granularity sampling detector in which the active medium is liquid argon (LAr) interspaced with layers of lead absorber.
The hadronic calorimeters are divided into three sections: a tile scintillator/steel calorimeter is used in both the barrel (\maths{\absEta<1.0}) and extended barrel cylinders (\maths{0.8<\absEta<1.7}) while the hadronic endcap (\maths{1.5<\absEta<3.2}) consists of LAr/copper calorimeter modules.
The forward calorimeter measures both electromagnetic and hadronic energy in the range \maths{3.2<\absEta<4.9} using LAr/copper and LAr/tungsten modules.

The MBTS system consists of 32 scintillator counters, organized into two disks with one on each side of the detector.
They are located in front of the end-cap calorimeter cryostats and cover the region \maths{2.1<\absEta<3.8}.

The online trigger selection used in this analysis employs the minimum bias and calorimeter jet triggers~\cite{Aad:2012xs}.
The minimum bias triggers are only available at the hardware level, while the calorimeter triggers have both hardware and software levels.
The Level-1 (L1) hardware-based trigger provides a fast, but low-granularity, reconstruction of energy deposited in towers in the calorimeter; the Level-2 (L2) software implements a simple jet reconstruction algorithm in a window around the region triggered at L1; and finally the Event Filter (EF) performs a more detailed jet reconstruction procedure taking information from the entirety of the detector.
The efficiency of jet triggers is determined using a bootstrap method, starting from the fully efficient MBTS trigger~\cite{Aad:2012xs}.
Between March and August of 2010, only L1 information was used to select events; both the L1 and L2 stages were used for the remainder of the 2010 data-taking period and all three levels were required for data taken during 2011.

\section{\MC event simulation}\label{sec:mc}
Simulated proton--proton collisions at \maths{\sqrt{s} = \unit{7}{\TeV}} were generated using the \Pythia 6.4~\cite{Sjostrand:2006za} program.
These were used only to derive systematic uncertainties and to correct for detector effects; for this purpose they are compared against uncorrected data.
Additional samples used to compare theoretical predictions to the data are described in \sectionRef{sec:theory}.

The \Pythia program implements leading-order (LO) QCD matrix elements for \maths{2\rightarrow2} processes followed by \pT-ordered parton showers and the Lund string hadronisation model.
The underlying event in \Pythia is modelled by multiple-parton interactions interleaved with the initial-state parton shower.

The events were generated using the MRST LO* parton distribution functions (PDFs)~\cite{Martin:2009iq,Sherstnev:2008dm}.
Samples which simulated the data-taking conditions during 2010 (2011) used version 6.423 (6.425) of the generator, together with the ATLAS AMBT1~\cite{Aad:2010ac} (Perugia2011~\cite{Skands:2010ak}) underlying event tune.
For the samples simulating the data-taking conditions from 2011, additional $pp$ collisions were overlaid onto the hard scatter in the correct proportions to replicate this effect in the data.
The final-state particles were passed through a detailed \GEANT~\cite{Agostinelli:2002hh} simulation of the ATLAS detector~\cite{Aad:2010ah} before being reconstructed using the same software used to process data.

\section{Jet reconstruction}\label{sec:jet-reconstruction}
The collision events selected by the ATLAS trigger system were fully reconstructed offline.
Energy deposits in the calorimeter left by electromagnetic and hadronic showers were calibrated to the electromagnetic (EM) scale.
\footnote{The electromagnetic scale is the basic calorimeter signal scale for the ATLAS calorimeters. It gives the correct response for the energy deposited in electromagnetic showers, but it does not correct for the different response for hadrons.}
Three-dimensional topological clusters (``topoclusters'')~\cite{Aharrouche:2010zz} were constructed from seed calorimeter cells according to an iterative procedure designed to suppress electronic noise~\cite{ATL-LARG-PUB-2009-001-2}.
Each of these was then treated as a massless particle with direction given by its energy-weighted barycentre.
The topoclusters were then passed as input to the \FastJet~\cite{Cacciari:2005hq} implementation of the \antikt jet algorithm~\cite{Cacciari:2008gp} with distance parameter \maths{R=0.6} and full four-momentum recombination.

The jets built by the \antikt algorithm were then calibrated in a multi-step procedure.
Additional energy arising from ``in-time pileup'' (simultaneous $pp$ collisions within a single bunch crossing) was subtracted using a correction derived from data.
Each event was required to have at least one primary vertex, reconstructed using two or more tracks, each with \maths{\pT > \unit{400}{\MeV}} and the primary vertex with the highest \maths{\sum \pTsq} of tracks associated with it was identified as the origin of the hard scatter.
The jet position was recalibrated to point to this identified hard scatter primary vertex, rather than the geometric centre of the detector.
A series of \pT- and \maths{\eta}-dependent energy correction factors derived from simulated events were used to correct for the response of the detector to jets.
For the data collected during 2011, additional calibration steps were applied.
Energy contributions, which were usually negative, coming from ``out-of-time pileup'' (residual electronic effects from previous $pp$ collisions) were corrected for using an offset correction derived using simulation.

A final \insitu calibration, using \maths{Z}+jet balance, \maths{\gamma}+jet balance and \multijet balance, was then applied to correct for residual differences in jet response between the simulation and data.
The calibration procedure is described in more detail elsewhere~\cite{Aad:2011he,Aad:2014bia}.

\section{Event selection}\label{sec:event-selection}
The measurements were performed using only the data from specific runs and run periods in which the detector, trigger and reconstructed physics objects satisfied data-quality selection criteria.
Beam background was rejected by requiring at least one primary vertex in each event while selection requirements were applied to the hard-scatter vertex to minimise contamination from pileup.
For data collected during 2010, the event was required to have only one primary vertex with five or more associated tracks; the proportion of such events was 93\% in the early low-luminosity runs, falling to 21\% in the high-luminosity runs at the end of the year.
For data collected during 2011, the hard-scatter vertex was required to have at least three associated tracks.
Jets arising from pileup were rejected using the jet vertex fraction (JVF).
The JVF takes all tracks matched to the jet of interest and measures the ratio of \maths{\sum \pT} from tracks which originated in the hard-scatter vertex to the \maths{\sum \pT} of all tracks matched to the jet.
For this analysis only jets with \maths{\text{JVF}>0.75} were used; all other jets were considered to have arisen from pileup and were therefore ignored.
Due to the limited coverage of the tracking detectors, the JVF is only available for jets satisfying \maths{\absRap < 2.4}, which limits the acceptance in rapidity for 2011 data.

Due to the high instantaneous luminosity reached by the LHC, only high-threshold jet triggers remained unprescaled throughout the data-taking period in question.
Jets with transverse momentum below the lowest-threshold unprescaled trigger were therefore only recorded using prescaled triggers.
For data collected during 2011, the average transverse momentum (\pTbar) of the \dijet{s} was used to determine the most appropriate trigger to use for each event.
Among all of the triggers determined to be fully efficient at the particular \pTbar in question, the one which had been least prescaled was selected: only events passing this trigger were considered.

For the data collected during 2010, it was necessary to combine triggers from the central region (\maths{\absEta \leq 3.2}) and the forward region (\maths{3.1 < \absEta \leq 4.9}) of the detector, due to the large \DeltaY span under consideration.
In each of these regions, efficiency curves as a function of jet transverse momentum were calculated for each trigger on a per-jet basis, rather than the per-event basis described above.
The triggers were ordered according to their prescales and a lookup table was created, showing the point at which each trigger reached a plateau of 99\% efficiency.

An appropriate trigger was chosen for each of the two leading (highest transverse momentum) jets in each event; this was the lowest prescale trigger to have reached its efficiency plateau at the relevant \pT and \absRap.
The \dijet event was then accepted if the event satisfied the trigger appropriate to the leading jet, the subleading jet or both.
This procedure maximised event acceptance, since the random factor inherent in the trigger prescale meant that some events could be accepted based on the properties of the subleading jet even when the appropriate trigger for the leading jet had not fired.
In order to combine overlapping triggers with different prescales, the procedure detailed in Ref.~\cite{Lendermann:2009ah} was followed.
In some less well-instrumented or malfunctioning regions of the detector, the per-jet trigger efficiency plateau occurred at less than the usual 99\% point.
This introduced a measurable trigger ineffiency, which was corrected for by weighting events containing jets in these regions by the inverse of the efficiency.

Jets were required to have transverse momentum \maths{\pT > \unit{20 (30)}{\GeV}} for data collected during 2010 (2011), thus ensuring that they remained in a region for which the jet energy scale had been evaluated (see \sectionRef{sec:jet-reconstruction}).
Jets were restricted in rapidity to \maths{\absRap < 4.4} for data collected in 2010, with a stricter requirement of \maths{\absRap < 2.4} applied in 2011 to ensure that the JVF could be determined for all jets.
The two leading jets satisfying these criteria were then identified as the \dijet system of interest.
The event was rejected if the transverse momentum of the leading jet was below \unit{60}{\GeV} or if that of the subleading jet was below \unit{50}{\GeV}.
For the data collected during 2011, a minimum rapidity separation, \maths{\DeltaY \geq 1}, was required to enhance the physics of interest.
The veto scale, \Qnought, was set to \maths{\pT > \unit{20 (30)}{\GeV}} for data collected in 2010 (2011).

Jet cleaning criteria~\cite{ATLAS-CONF-2010-038,ATLAS-CONF-2012-020} were developed in order to reject fake jets, those which come from cosmic rays, beam halo or detector noise.
These criteria also removed jets which were badly measured due to falling into poorly instrumented regions.
Events collected in 2010 (2011) were rejected if they contained any jet with transverse momentum \maths{\pT > \unit{20 (30)}{\GeV}} that failed these cleaning cuts.
This requirement was also applied to the simulated samples where appropriate.

Additionally, a problem developed in the LAr calorimeter during 2011 running, resulting in a region in which energies were not properly recorded.
As a result, a veto was applied to events that had at least one jet with \maths{\pT > \unit{30}{\GeV}} falling in the vicinity of this region during the affected data-taking periods.
This effect was replicated in the relevant simulation samples, which were reweighted to the data to ensure that an identical proportion of such events were included.

\begin{figure*}[htpb]
  \centering
  \subfloat[]{
    \includegraphics[width=\paperdoublefig]{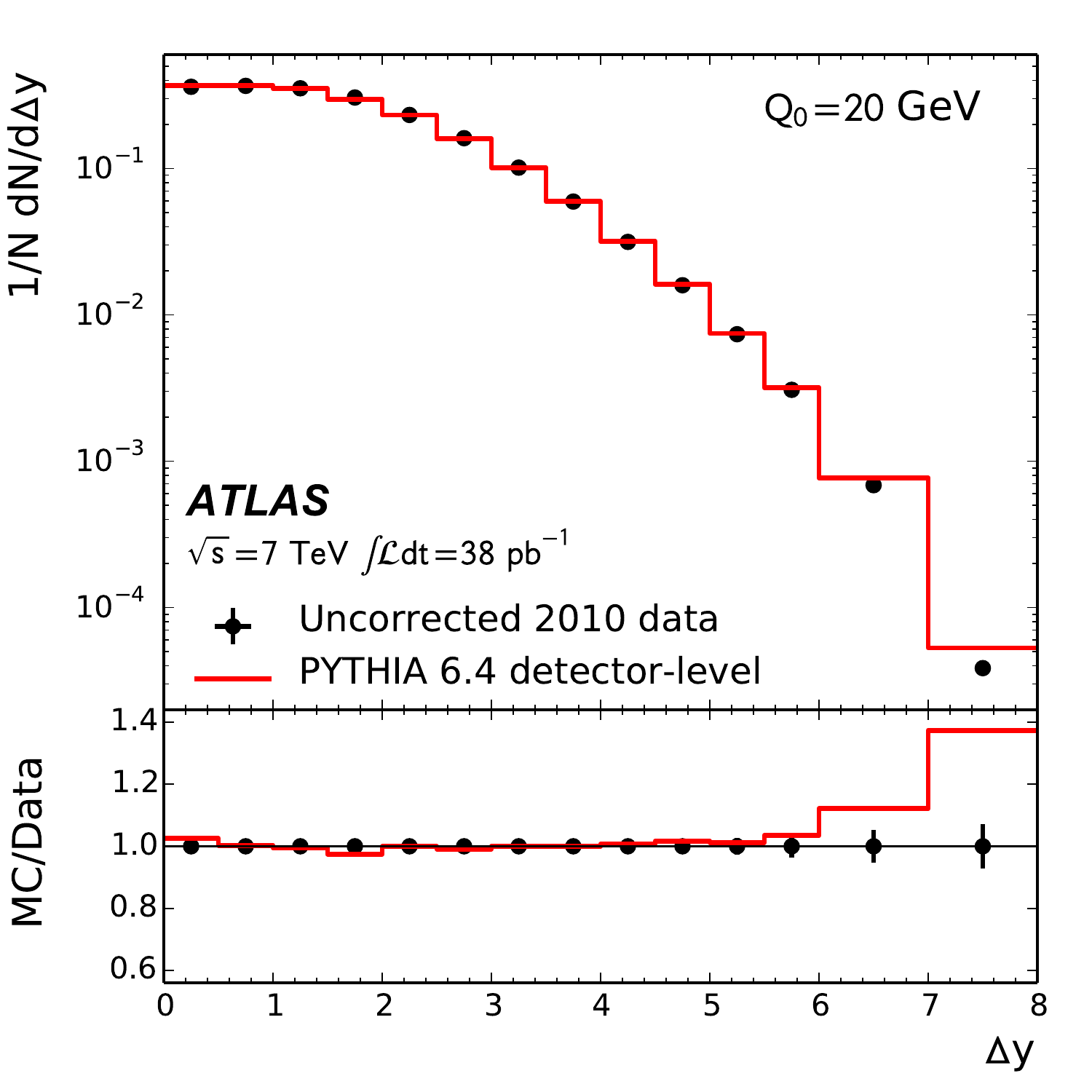}
    \label{fig:event-selection:Inclusive_dyBins}}
  \qquad
  \subfloat[]{
    \includegraphics[width=\paperdoublefig]{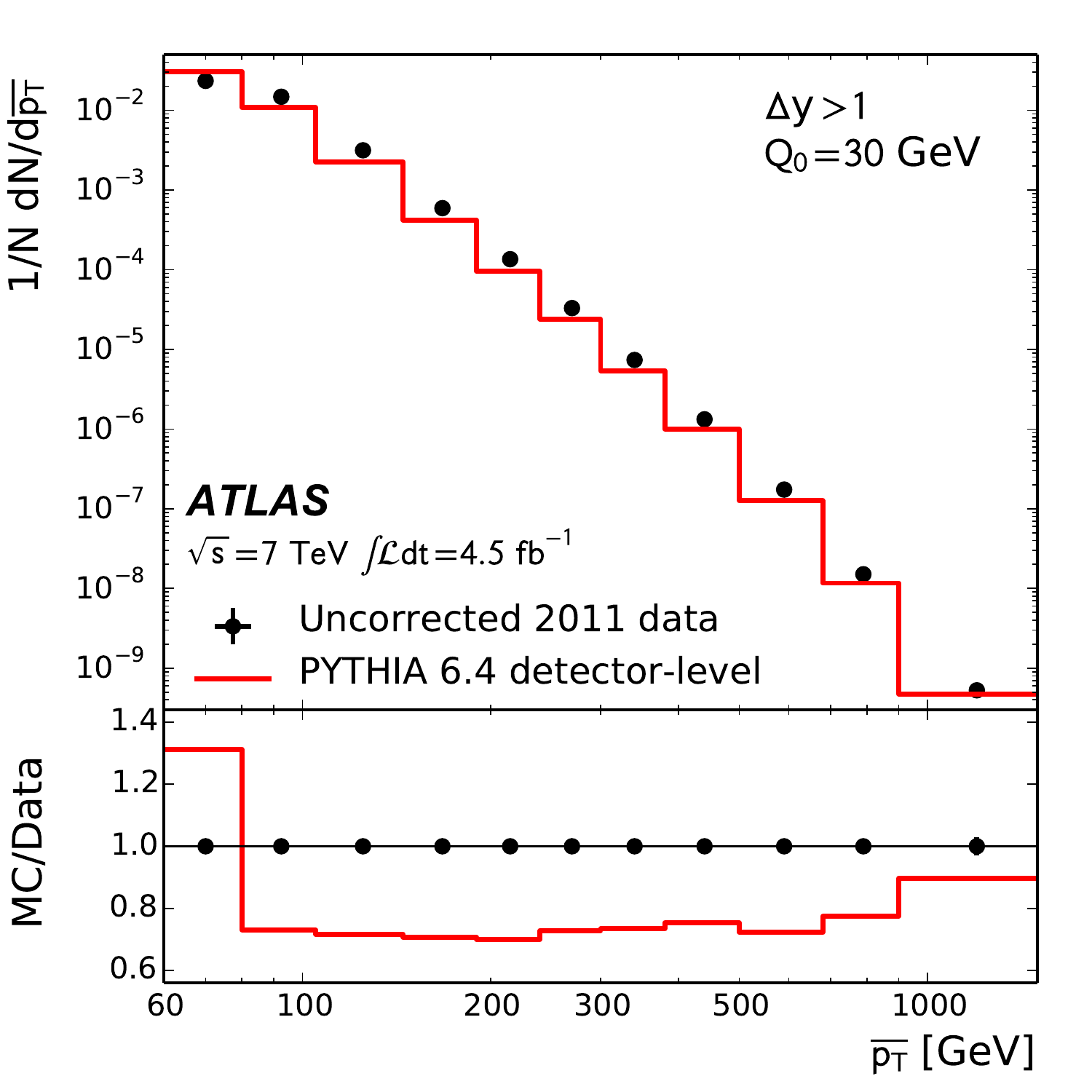}
    \label{fig:event-selection:Inclusive_pTbarBins}}
  \caption{Comparison between uncorrected data (black points) and detector level \Pythia 6.4 \MC events (solid line). Statistical errors on the data are shown. The normalised distribution of events is presented as a function of \protect\subref{fig:event-selection:Inclusive_dyBins} \DeltaY and \protect\subref{fig:event-selection:Inclusive_pTbarBins} \pTbar. The ratio of the \Pythia 6.4 prediction to the data is shown in the bottom panel.}
  \label{fig:event-selection:control}
\end{figure*}

In total, 1188583 events were accepted from the data collected in 2010, with 852030 of these being gap events: those with no additional jets above the veto scale in the rapidity interval between the \dijet{s}.
For data collected during 2011, 1411676 events were accepted, with 938086 of these being gap events.
Data from 2010 and from 2011 were compared after being analysed separately and were found to agree to within the experimental uncertainties in a region kinematically accessible with both datasets.
\FigureRef{fig:event-selection:control} shows the comparison between detector level data and \Pythia 6.4 simulation in \dijet events.
The normalised number of events is presented as a function of \DeltaY in \figureSubRef{fig:event-selection:Inclusive_dyBins} and of \pTbar in \figureSubRef{fig:event-selection:Inclusive_pTbarBins}.
In both cases, and in all similar distributions, the \Pythia 6.4 event generator and \GEANT detector simulation give a fair description of the uncorrected data.

\section{Correction for detector effects}\label{sec:unfolding}
Before comparing to theoretical predictions, the data are corrected for all experimental effects so that they correspond to the particle-level final state.
This comprises all stable particles, defined as those with a proper lifetime longer than \unit{10}{\pico\second}, including muons and neutrinos from decaying hadrons~\cite{Buttar:2008jx}.
The correction for detector resolutions and inefficiencies is made by unfolding the measured distributions using the Bayesian procedure~\cite{DAgostini:2010:BayesianUnfolding} implemented in the \RooUnfold framework~\cite{Adye:2011gm}.

Bayesian unfolding entails using simulated events to calculate a transfer matrix which encodes bin-to-bin migrations between particle-level distributions and the equivalent reconstructed distributions at detector level.
A series of bin transition probabilities are obtained from the matrix, and Bayes' theorem is used to calculate the corresponding inverse probabilities; the process is then repeated iteratively.
In this paper, the unfolding is performed using the \Pythia 6.4 samples described in \sectionRef{sec:mc} and the number of iterations is set to two throughout, as this was found to be sufficient to achieve convergence.

As the results shown here are constructed from multi-dimensional distributions, this must also be taken into account when unfolding.
Each distribution is unfolded in two or three dimensions, with these dimensions being relevant combinations of \DeltaPhi, \DeltaY, \pTbar, \CosDeltaPhi, \CosTwoDeltaPhi and the classification of the event as gap or non-gap.
This means that the transfer matrices are four-dimensional or six-dimensional, rather than the usual two-dimensional case.
This allows the effect of all possible bin migrations to be evaluated.

The statistical uncertainties are estimated by performing pseudo-experiments~\cite{Efron:1979}.
Each event in data is assigned a weight drawn from a Poisson distribution with unit mean for each pseudo-experiment and these weighted events are used to build a series of one thousand replicas for each distribution.
Each of these replicas is unfolded, and the root-mean-squared spread around the nominal value is used to measure the statistical error on the unfolded result.

Possible bias arising from mismodelling of the distributions considered here is evaluated by performing a self-consistency check using \Pythia 6.4 events.
The \Pythia 6.4 simulation is reweighted on an event-by-event basis using a three-dimensional function which is chosen in such a way as to ensure that the output of this reweighting step will approximate the uncorrected detector level data.
For the \Pythia sample simulated to replicate 2010 data-taking conditions, the reweighting is carried out as a function of \DeltaY, \DeltaPhi and the highest \pT among jets in the rapidity interval; for the sample simulated to replicate 2011 data-taking conditions, \pTbar is used instead of \DeltaY.
This reweighted detector level \Pythia 6.4 sample is then unfolded using the original transfer matrix and the result is compared with the particle-level spectrum, which was itself implicitly modified through the event-by-event reweighting.
Any remaining difference between these distributions is then taken as a systematic uncertainty associated with the unfolding procedure.

\section{Systematic uncertainties}\label{sec:systematics}
For the most part, the dominant systematic uncertainty on these measurements is the one coming from the jet energy scale (JES) calibration.
This uncertainty was determined using a combination of \insitu calibration techniques, as detailed earlier, test-beam data and \MC modelling~\cite{Aad:2011he,Aad:2014bia}.
It comprises 13 independent components for data taken in 2010 and 64 components for the 2011 data.
The uncertainty components fully account for the differences in jet calibration discussed in \sectionRef{sec:jet-reconstruction}, thus ensuring that data collected in the two years are fully compatible within their uncertainties.

For each component, all jet energies and transverse momenta are shifted up or down by one standard deviation of the uncertainty and the shifted jets are then passed through the full analysis chain.
The measured distributions are unfolded and compared to the nominal distribution; the difference between these is taken as the uncertainty for the component in question.
These fractional differences are then combined in quadrature, since the components are uncorrelated, to compute the jet energy scale uncertainty.

The uncertainty on the energy resolution of jets (JER) is derived \insitu, using \dijet balance techniques and the bisector method; it is then cross-checked through comparison with simulation~\cite{Aad:2012ag}.
Jet angular resolutions are estimated using simulated events and cross-checked using \insitu techniques, where good agreement is observed with the simulation.
These resolution uncertainties are propagated through the unfolding procedure by smearing the energy or angle of each reconstructed jet in each simulated event by a Gaussian function, with its width given by the quadratic difference between the nominal resolution and the resolution after shifting by the resolution uncertainty.
This procedure is repeated one thousand times for each jet, to remove the effects of statistical fluctuations.
The resulting smeared events are used to calculate a modified transfer matrix incorporating the resolution uncertainty; this matrix is then used to unfold the data.
The ratio of this distribution to the distribution unfolded using the nominal transfer matrix is taken as a systematic uncertainty.

The trigger efficiency correction which is applied to events with jets falling into poorly measured detector regions also has an associated systematic uncertainty.
This is determined by increasing or decreasing the measured inefficiencies by an absolute shift of 10\%, with a maximum efficiency of 100\%.
The full correction procedure is then carried out using these new correction factors and the difference between this and the nominal distribution is taken as a systematic uncertainty on the measurement.

The effect of statistical fluctuations in the samples used to derive these uncertainties is also estimated by performing pseudo-experiments.
Each event in the sample is assigned a weight drawn from a Poisson distribution with unit mean for each pseudo-experiment and these weighted events are used to build a series of one thousand replicas of the transfer matrices.
These replicas are then used to unfold the nominal data sample; the root-mean-squared spread around the nominal value provides the systematic error due to the limited statistical precision of the \MC samples used.

For each of the systematic variations considered, the pseudo-experiment approach applied to the data is used to determine statistical uncertainties on each distribution and correlations between bins.
Each fractional uncertainty was smoothed to remove these statistical fluctuations before the uncertainties were combined.
To do this, each systematic component was rebinned until each bin showed a statistically significant deviation from the nominal value.
This rebinned distribution was then smoothed using a Gaussian kernel and the smoothed function was evaluated at each of the original set of bin centres.
The overall fractional uncertainty was then obtained by summing the individual sources in quadrature.

\begin{figure*}[htpb]
  \centering
  \subfloat[]{
    \includegraphics[width=\paperdoublefig]{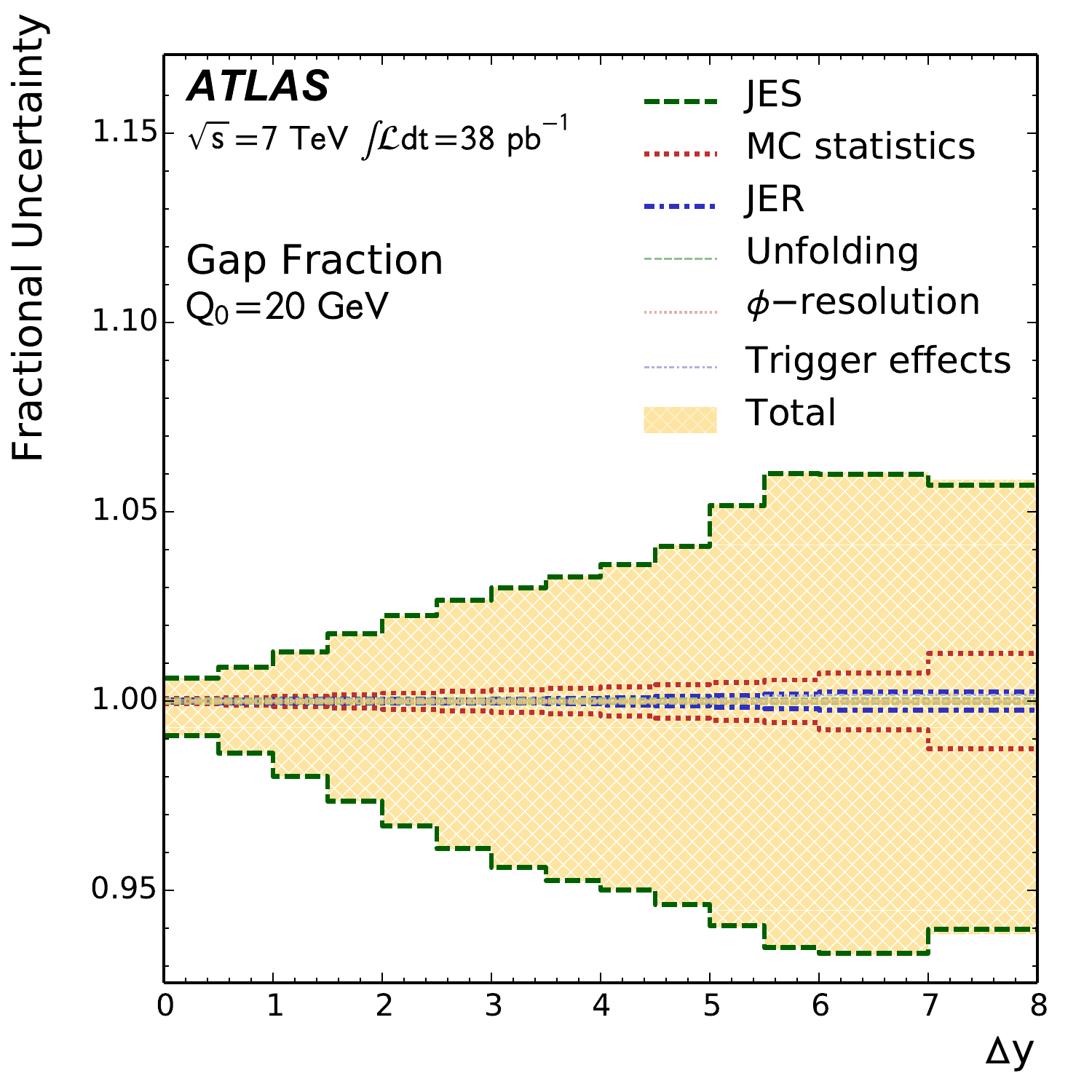}
    \label{fig:syst:total_gf_dy}}
  \qquad
  \subfloat[]{
    \includegraphics[width=\paperdoublefig]{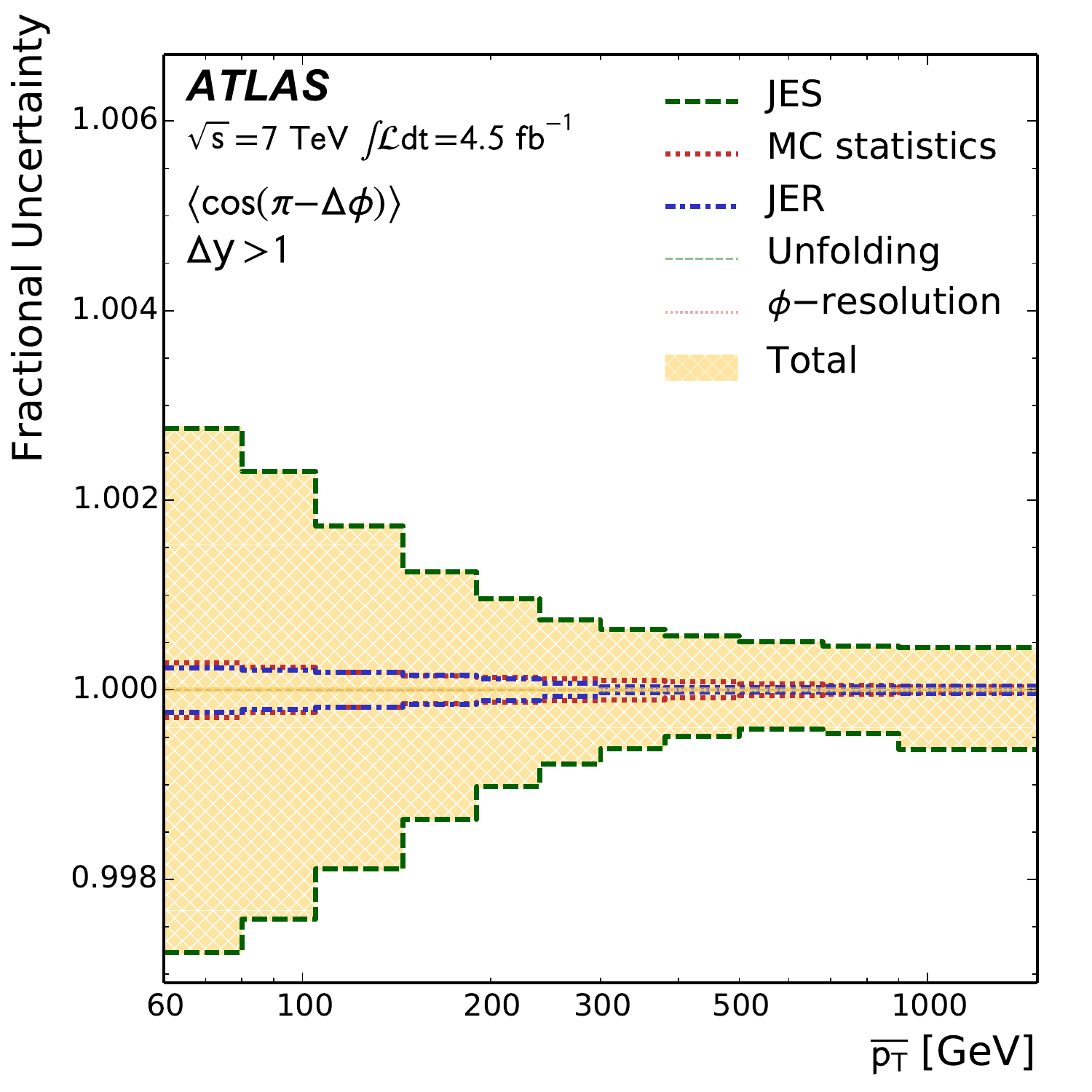}
    \label{fig:syst:total_CosDeltaPhi_pTbar_incl}}
  \caption{Summary of systematic uncertainties on \protect\subref{fig:syst:total_gf_dy} the gap fraction as a function of \DeltaY and \protect\subref{fig:syst:total_CosDeltaPhi_pTbar_incl} the \meanCosDeltaPhi distribution as a function of \pTbar. Here the systematic uncertainties from the jet energy scale (dark dashes), \MC statistical precision (dark dots), jet energy resolution as (dark dashed-dots), unfolding (light dashes), jet \maths{\phi} resolution (light dots) and from residual trigger inefficiencies (light dashed-dots) are shown. The total systematic uncertainty (light cross-hatched area) is also shown.}
  \label{fig:syst:total}
\end{figure*}

The uncertainty on the unfolding procedure, estimated as described in \sectionRef{sec:unfolding} is also important in some distributions.
Other sources of uncertainty, such as residual pileup effects, were examined, found to be negligible and therefore ignored.
There is also an additional uncertainty on the luminosity calibration which is not included here.
For the \xs{s} this is $1.8\%$ while it cancels for all other distributions.
\FigureRef{fig:syst:total} shows the summary of systematic uncertainties for two sample distributions: \figureSubRef{fig:syst:total_gf_dy} for the gap fraction as a function of \DeltaY and \figureSubRef{fig:syst:total_CosDeltaPhi_pTbar_incl} for the \meanCosDeltaPhi distribution as a function of \pTbar.

\section{Theoretical predictions}\label{sec:theory}
Two state-of-the-art theoretical predictions, namely \HEJ (High Energy Jets)~\cite{Andersen:2010ih,Andersen:2011hs} and the \Powheg BOX~\cite{Nason:2004rx,Frixione:2007vw,Alioli:2010xd}, are considered in this paper.

\HEJ provides a leading-logarithmic (LL) calculation of the perturbative terms that dominate the production of \multijet events when the jets span a large range in rapidity~\cite{Andersen:2009nu,Andersen:2011hs,Andersen:2009he}.
This formalism resums logarithms relevant in the Mueller-Navelet~\cite{Mueller:1986ey} limit, and incorporates a contribution from all final states with at least two hard jets.
The purely partonic multi-jet output from \HEJ can also be interfaced to the \Ariadne parton shower framework~\cite{Lonnblad:1992tz} to evolve the prediction to the hadron-level final state~\cite{Andersen:2011zd}.
The \Ariadne program is based on the colour-dipole cascade model~\cite{Lonnblad:1994wk} in which gluon emissions are modelled as radiation from colour-connected partons, and provides soft and collinear radiation down to the hadronic scale, using \Pythia 6.4 for hadronisation.
This accounts for radiation in the rapidity region outside the \multijet system modelled by \HEJ and represents a contribution from small-$x$, BFKL-like, logarithmic terms.

The \Powheg Box (version r2169) provides a full next-to-leading-order \dijet calculation and is interfaced to either \PythiaEight~\cite{Sjostrand:2007gs} (AU2 tune with $\alpha_\text{s}$ matching for the ISR~\cite{ATL-PHYS-PUB-2012-003}) or \Herwig~\cite{Corcella:2000bw} (AUET1 tune~\cite{ATL-PHYS-PUB-2010-014}) to provide all-order resummation of soft and collinear emissions using the parton shower approximation.
The advantage over the simple \maths{2\rightarrow2} matrix elements is that the emission of an additional third hard parton is calculated exactly in pQCD, allowing observables that depend on the third jet to be calculated with good accuracy~\cite{Alioli:2010xa}.

The prediction provided by \Powheg uses the DGLAP formalism, while that provided by \HEJ is based on BFKL.
For all theoretical predictions, events were generated using the CT10 PDF set~\cite{Pumplin:2002vw}.
The orthogonal error sets provided as part of the CT10 PDF set were used in order to evaluate the uncertainty inherent in the PDF, at the 68\% confidence level, following the CTEQ prescription~\cite{Campbell:2006wx}.
The default choice for the renormalisation and factorisation scales, which are constrained to be identical in \HEJ, was the transverse momentum of the leading parton in each event.
The uncertainty due to higher-order corrections was estimated, for both the \Powheg and \HEJ predictions, by increasing and decreasing the scale by a factor of two and taking the envelope of these variations.

The scale uncertainty and PDF uncertainties were combined in quadrature to construct an overall uncertainty for each prediction.
The PDF uncertainties are small across all of the phase-space regions considered in this paper and hence the predominant contribution to the uncertainty comes from the scale uncertainty.
Theoretical uncertainties on the \HEJA prediction are not currently calculable and are not shown here.
The range covered by the central values of the \PowhegP and \PowhegH predictions gives an estimate of the uncertainty inherent in the parton shower matching procedure.
This range, together with the uncertainty band on the \PowhegP prediction, can be considered together as a total theoretical uncertainty on the NLO+DGLAP prediction that can be compared to the predictions given by \HEJ and \HEJA.

Finally, in order to allow comparisons against the fully corrected data distributions presented in this paper, the partons from \HEJ or the final-state particles from \HEJA, \PowhegP and \PowhegH were clustered together using the same jet algorithm and parameters as for the data.

\section{Results and discussion}\label{sec:results}
The fully corrected data are compared to next-to-leading-order theoretical predictions from \Powheg and \HEJ, as explained in \sectionRef{sec:theory}.
The \Powheg prediction is presented after parton showering, hadronisation and underlying event simulation with either \PythiaEight or \Herwig.
As the uncertainties on these two \Powheg predictions are highly correlated, uncertainties are only shown on the \PowhegP prediction, with only the central value of the \PowhegH prediction presented.
Two \HEJ curves are presented: one a pure parton-level prediction and the second after interfacing with the \Ariadne parton shower.
The central value of the \HEJA prediction is shown, together with the statistical uncertainty on this prediction.

\subsection{Gap fraction and mean jet multiplicity}
\begin{figure*}[htpb]
  \centering
  \subfloat[]{
    \includegraphics[width=\paperdoublefig]{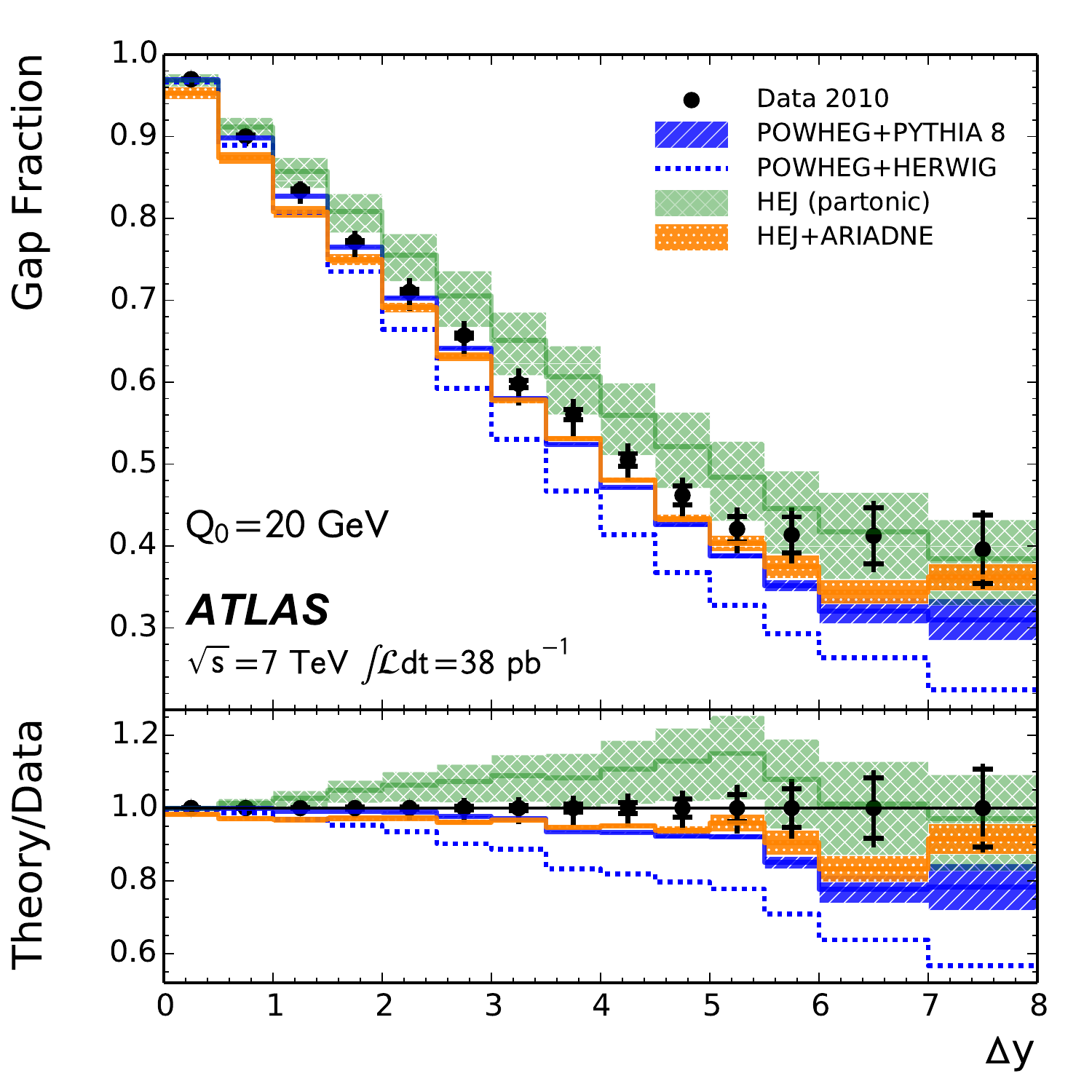}
    \label{fig:results:GapFraction_dyBins}}
  \qquad
  \subfloat[]{
    \includegraphics[width=\paperdoublefig]{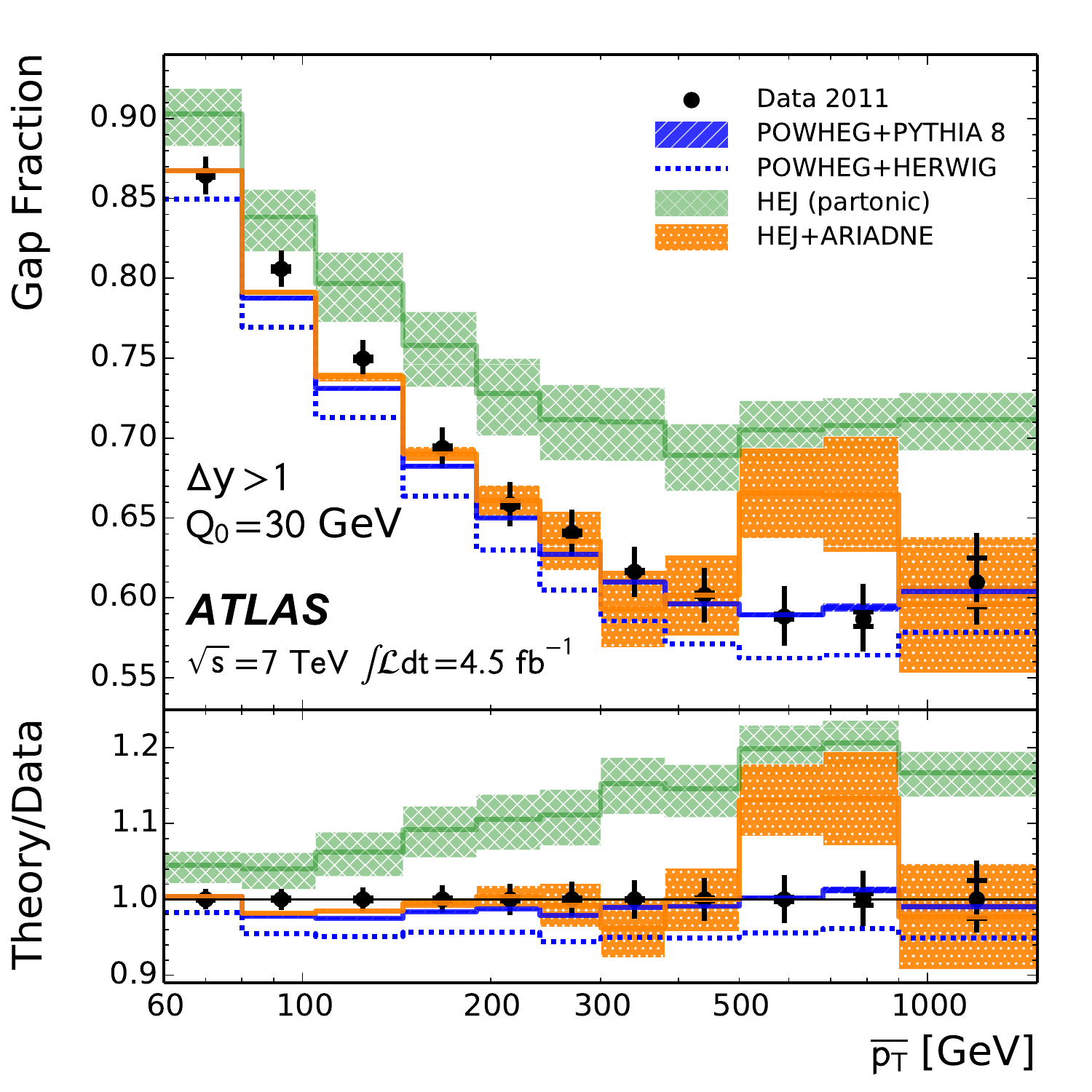}
    \label{fig:results:GapFraction_pTbarBins}}
  \caption{The measured gap fraction (black dots) as a function of \protect\subref{fig:results:GapFraction_dyBins} \DeltaY and \protect\subref{fig:results:GapFraction_pTbarBins} \pTbar. The inner error bars represent statistical uncertainty while the outer error bars represent the quadrature sum of the systematic and statistical uncertainties. For comparison, the predictions from parton-level \HEJ (light-shaded cross-hatched band), \HEJA (mid-shaded dotted band), \PowhegP (dark-shaded hatched band) and \PowhegH (dotted line) are also included. The ratio of the theory predictions to the data is shown in the bottom panel.}
  \label{fig:results:GapFraction}
\end{figure*}

\begin{figure*}[htpb]
  \centering
  \subfloat[]{
    \includegraphics[width=\paperdoublefig]{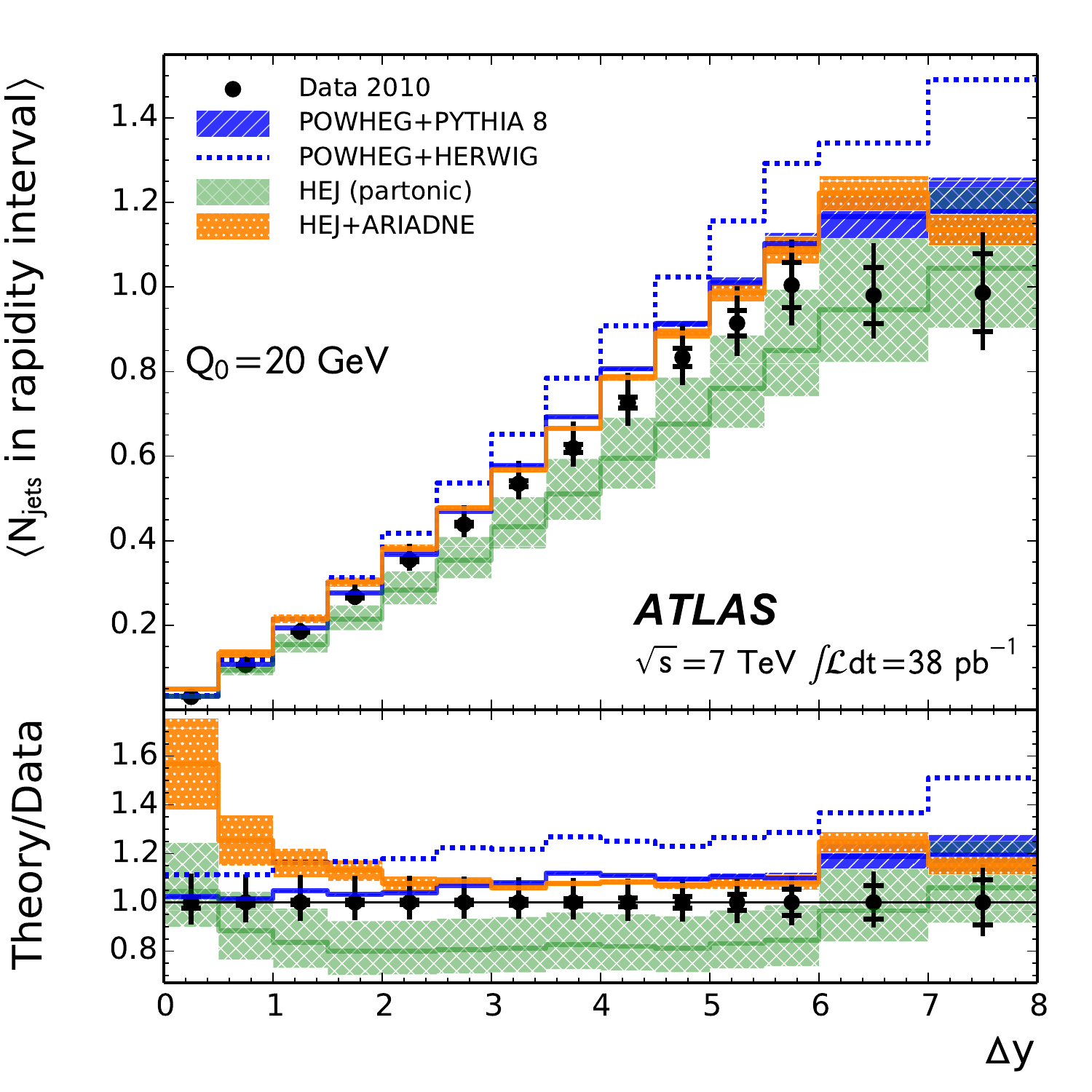}
    \label{fig:results:nGapJets_dyBins}}
  \qquad
  \subfloat[]{
    \includegraphics[width=\paperdoublefig]{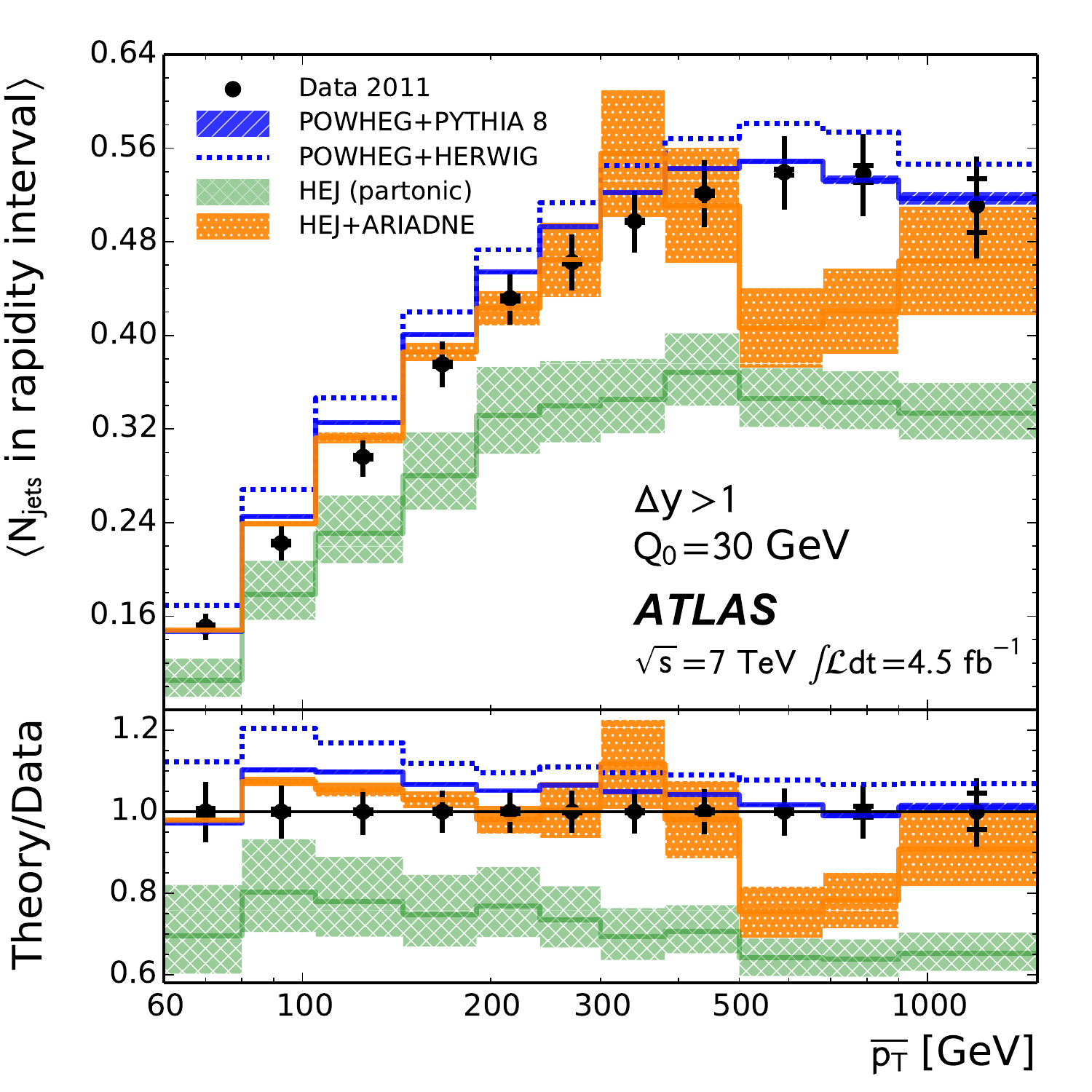}
    \label{fig:results:nGapJets_pTbarBins}}
  \caption{The mean number of jets above the veto threshold in the rapidity interval bounded by the \dijet system measured in data as a function of \protect\subref{fig:results:nGapJets_dyBins} \DeltaY and \protect\subref{fig:results:nGapJets_pTbarBins} \pTbar. For comparison, the \HEJ, \HEJA, \PowhegP and \PowhegH predictions are presented in the same way as \figureRef{fig:results:GapFraction}.}
  \label{fig:results:nGapJets}
\end{figure*}

\FigureRefTwo{fig:results:GapFraction}{fig:results:nGapJets} show the gap fraction and the number of jets in the rapidity gap, respectively, as functions of \DeltaY and \pTbar.
Na\"ively, it is expected from pQCD that the number of events passing the jet veto should be exponentially suppressed as a function of \DeltaY and \lnpTbarQnought due to the exchange of colour in the \maths{t}-channel~\cite{Forshaw:2009fz}.
However, non-exponential behaviour may become apparent in the tails of these distributions as the steeply falling parton distribution functions can reduce the probability of additional quark and gluon radiation from the \dijet system and increase the gap fraction~\cite{DuranDelgado:2011tp}.
This can be understood by considering the behaviour at extreme values of \DeltaY or \pTbar, when all of the collision energy is used to create the \dijet pair and little is available for additional radiation.
As the gap fraction is expected to be smooth it must therefore begin increasing at some point, so as to reach unity when this kinematic limit is obtained.
Furthermore, since the \xs for QCD colour-singlet exchange increases with jet separation~\cite{Mueller:1992pe}, any contribution from such processes would also lead to an increase in the gap fraction at large \DeltaY.

The data do indeed show exponential behaviour in \figureRef{fig:results:GapFraction} at low values of \DeltaY and \pTbar, but deviate from purely exponential behaviour at the highest values of \DeltaY and \pTbar, with the gap fraction reaching a plateau in both distributions.
For the \pTbar distribution, this plateau is qualitatively reproduced by all the predictions considered here, even those which do not provide good overall agreement with the data.
The plateau observed in data for the \DeltaY distribution is not, however, as prominent in any of the theoretical predictions, which all continue to fall as \DeltaY increases.
A similar excess was observed in previous experiments~\cite{Derrick:1995pb,Adloff:2002em,Chekanov:2006pw,Abachi:1994hb,Abe:1994de,Abe:1997ie,Abe:1998ip,Abbott:1998jb} and was attributed to colour-singlet exchange effects.
However, here the spread of theoretical predictions is too large to allow definite conclusions to be drawn and improved calculations are needed before a quantitative statement can be made.

In the high-\DeltaY region, both \Powheg predictions slightly underestimate the gap fraction and hence overestimate the mean jet multiplicity in the rapidity interval.
Partonic \HEJ slightly overestimates the gap fraction for intermediate values of \DeltaY.
Interfacing \HEJ to \Ariadne improves the description of the data across the \DeltaY spectrum.

\PowhegP, which resums soft and collinear emissions through the parton shower approximation, provides a good description of the gap fraction and the mean jet multiplicity distributions as a function of \pTbar.
On the other hand, the \PowhegH model, which also provides a similar resummation, consistently predicts too much jet activity across the \pTbar range.
Conversely, \HEJ, which does not attempt to resum these soft and collinear terms, provides a poor description of the data in the large \lnpTbarQnought limit.
Significantly improved agreement with the data is seen when interfacing \HEJ to the \Ariadne parton shower model, which performs a resummation of these terms.
In fact, the prediction from \HEJA is similar to that from \PowhegP for most values of \DeltaY and \pTbar.

\subsection{Azimuthal decorrelations}
\begin{figure*}[htpb]
  \centering
  \subfloat[]{
    \includegraphics[width=\paperdoublefig]{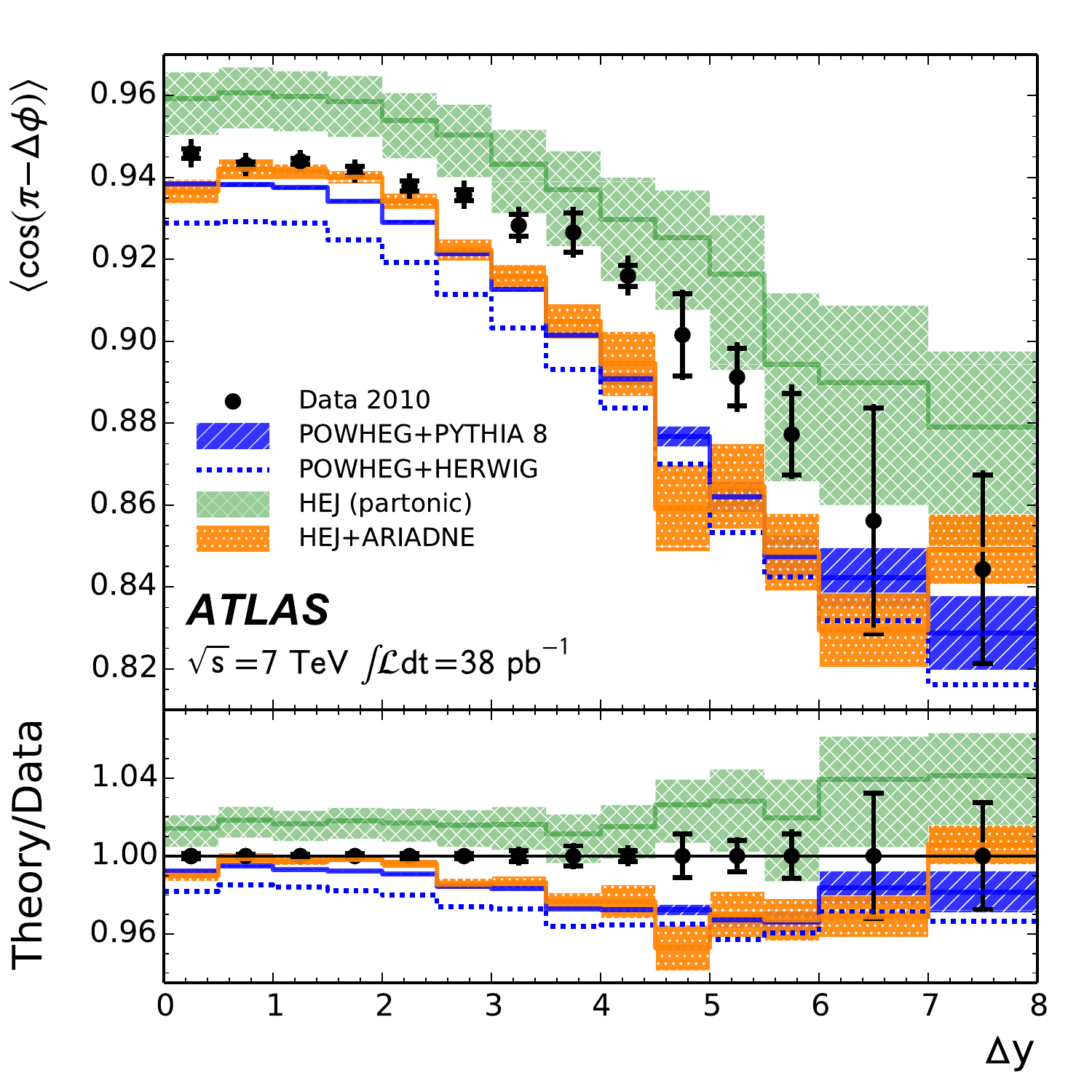}
    \label{fig:results:CosDeltaPhi_Inclusive_dyBins}}
  \qquad
  \subfloat[]{
    \includegraphics[width=\paperdoublefig]{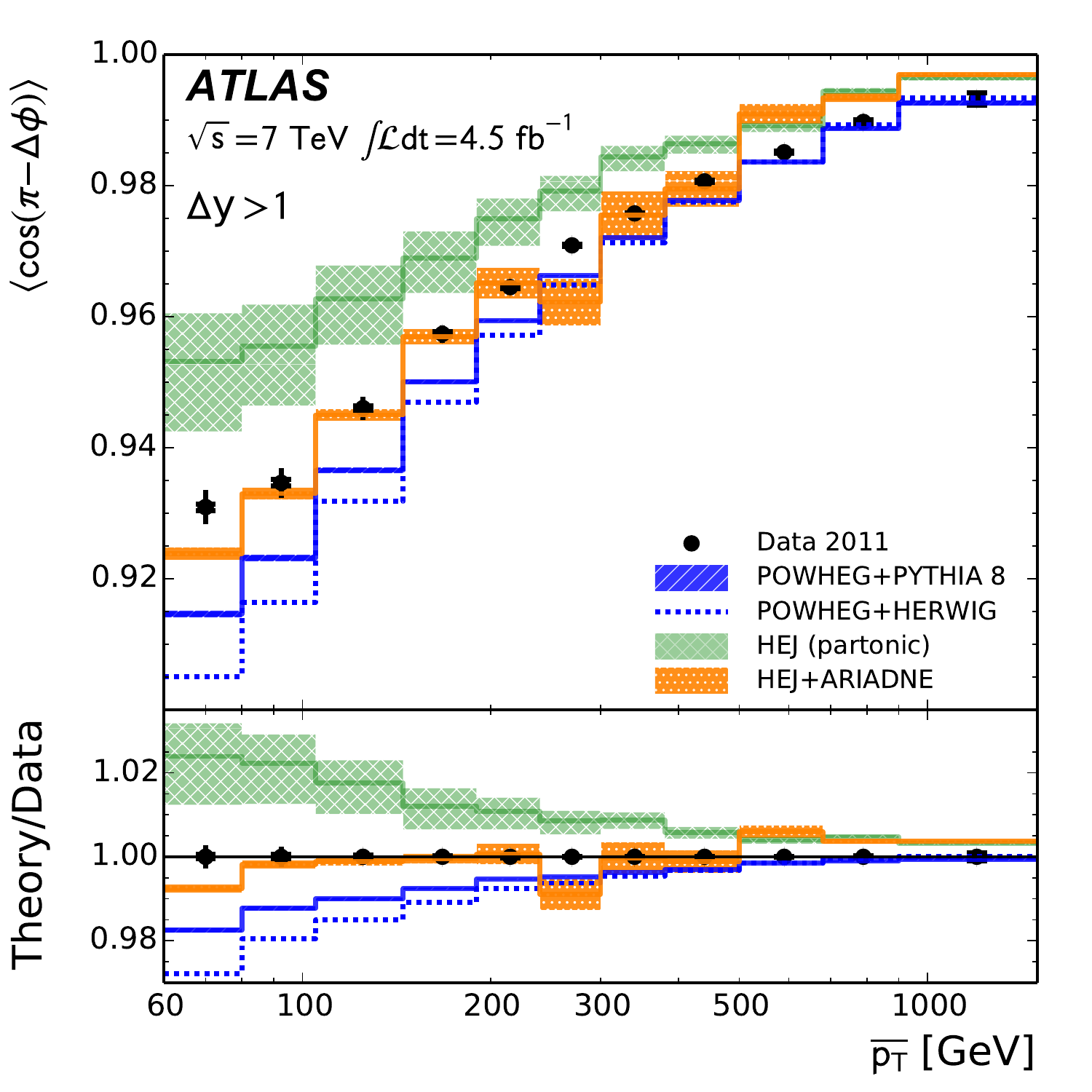}
    \label{fig:results:CosDeltaPhi_Inclusive_pTbarBins}}
  \\
  \subfloat[]{
    \includegraphics[width=\paperdoublefig]{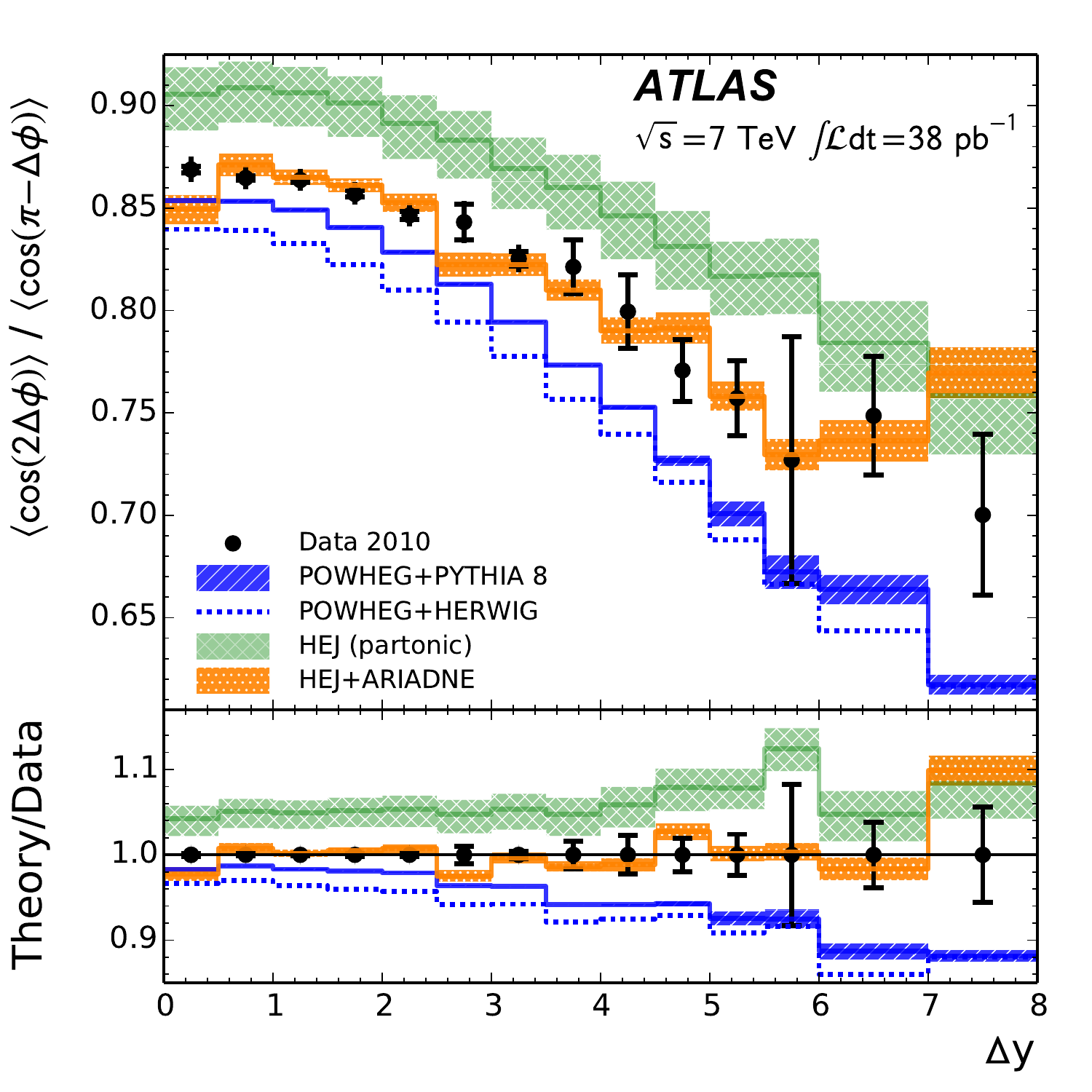}
    \label{fig:results:C2C1_Inclusive_dyBins}}
  \qquad
  \subfloat[]{
    \includegraphics[width=\paperdoublefig]{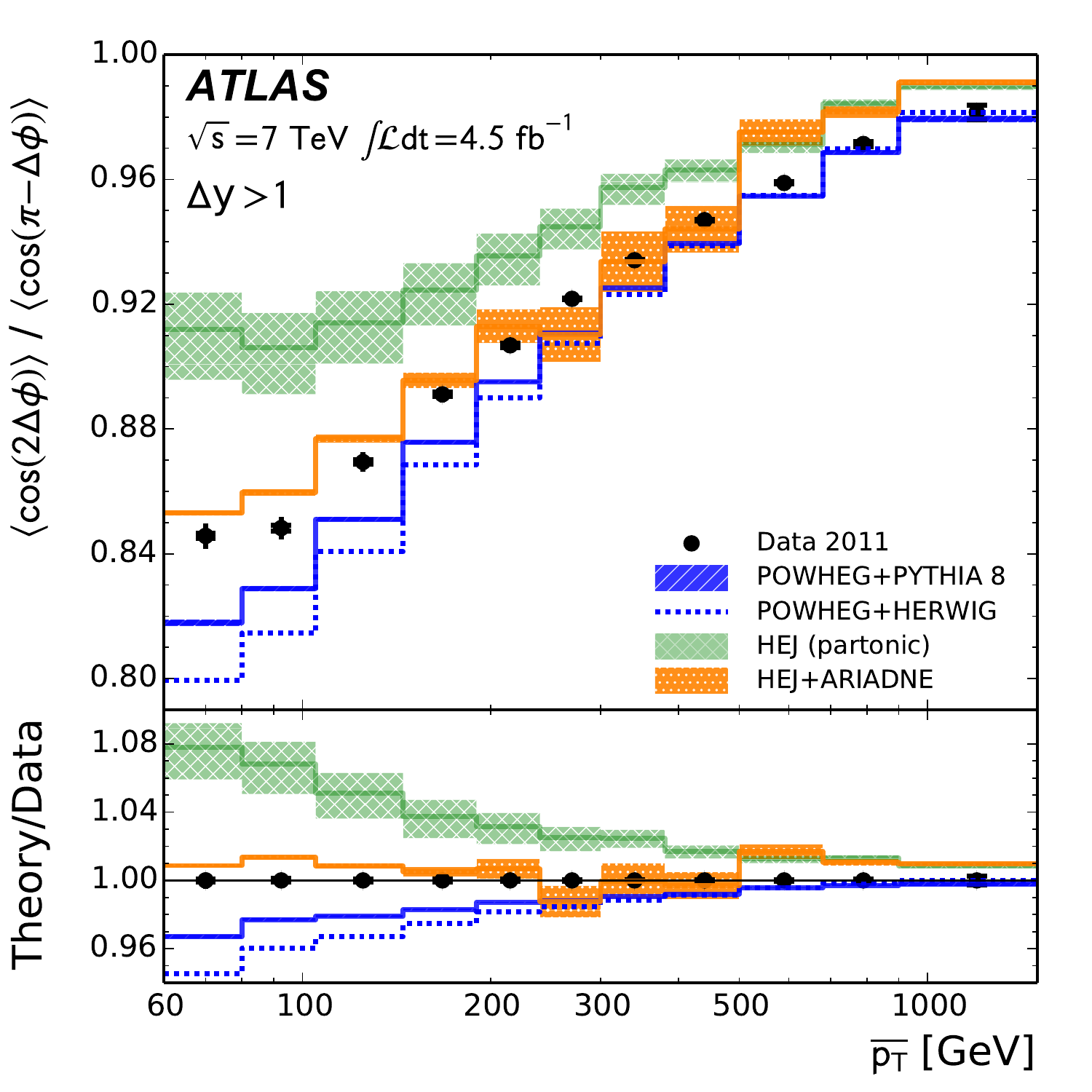}
    \label{fig:results:C2C1_Inclusive_pTbarBins}}
  \caption{The measured \protect\subref{fig:results:CosDeltaPhi_Inclusive_dyBins}\protect\subref{fig:results:CosDeltaPhi_Inclusive_pTbarBins} \meanCosDeltaPhi and \protect\subref{fig:results:C2C1_Inclusive_dyBins}\protect\subref{fig:results:C2C1_Inclusive_pTbarBins} \CtwoCone distributions as a function of \protect\subref{fig:results:CosDeltaPhi_Inclusive_dyBins}\protect\subref{fig:results:C2C1_Inclusive_dyBins} \DeltaY and \protect\subref{fig:results:CosDeltaPhi_Inclusive_pTbarBins}\protect\subref{fig:results:C2C1_Inclusive_pTbarBins} \pTbar. For comparison, the \HEJ, \HEJA, \PowhegP and \PowhegH predictions are presented in the same way as \figureRef{fig:results:GapFraction}.}
  \label{fig:results:CosDeltaPhi_C2C1_Inclusive}
\end{figure*}

\FigureRef{fig:results:CosDeltaPhi_C2C1_Inclusive} shows the \meanCosDeltaPhi and \CtwoCone distributions, as functions of \DeltaY and \pTbar, for inclusive \dijet events.
For the azimuthal moments, \meanCosNDeltaPhi, a decrease (increase) in azimuthal correlation manifests as a decrease (increase) in the azimuthal moment.
As the \dijet{s} deviate from a back-to-back topology, the second azimuthal moment falls more rapidly than the first (in the region \maths{\DeltaPhi > \pi/2} where the majority of events lie).
The ratio \CtwoCone is, therefore, expected to show a similar, but more pronounced, dependence on azimuthal correlation to that seen in the moments.

The data show the expected qualitative behaviour of a decrease in azimuthal correlation with increasing \DeltaY and increase in azimuthal correlation with increasing \pTbar.
Both \Powheg predictions underestimate the degree of azimuthal correlation except in the high-\pTbar region, while \HEJ predicts too much azimuthal correlation.
In both cases, the changing degree of correlation with \DeltaY and \pTbar is, for the most part, well described by the predictions.
The largest differences between the predictions and the data are seen at high \DeltaY and low \pTbar.

Additionally, it can be seen that the separation between theoretical predictions for the ratio \CtwoCone is significantly greater than for the \meanCosDeltaPhi distribution alone, considering the uncertainties of these predictions.
This means that the ratio gives enhanced discrimination between the DGLAP-like \Powheg and BFKL-like \HEJ predictions, as predicted by theoretical calculations~\cite{Colferai:2010wu}.
Here, neither \HEJ nor \Powheg provide good agreement with the data.
However, the \HEJA prediction gives a good description of the data for both low-\pTbar and for \DeltaY.

\begin{figure*}[htpb]
  \centering
  \subfloat[]{
    \includegraphics[width=\paperdoublefig]{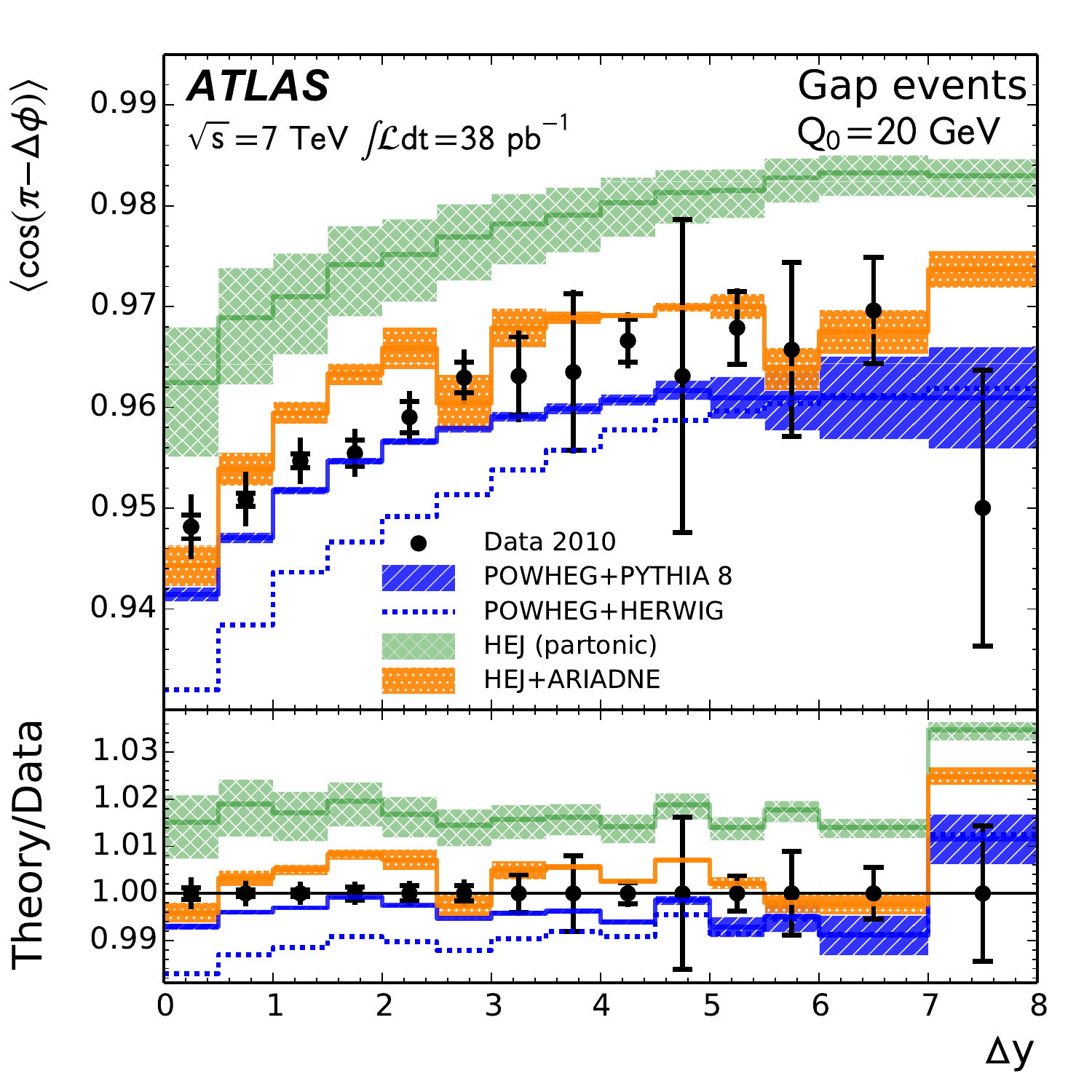}
    \label{fig:results:CosDeltaPhi_Gap_dyBins}}
   \qquad
  \subfloat[]{
    \includegraphics[width=\paperdoublefig]{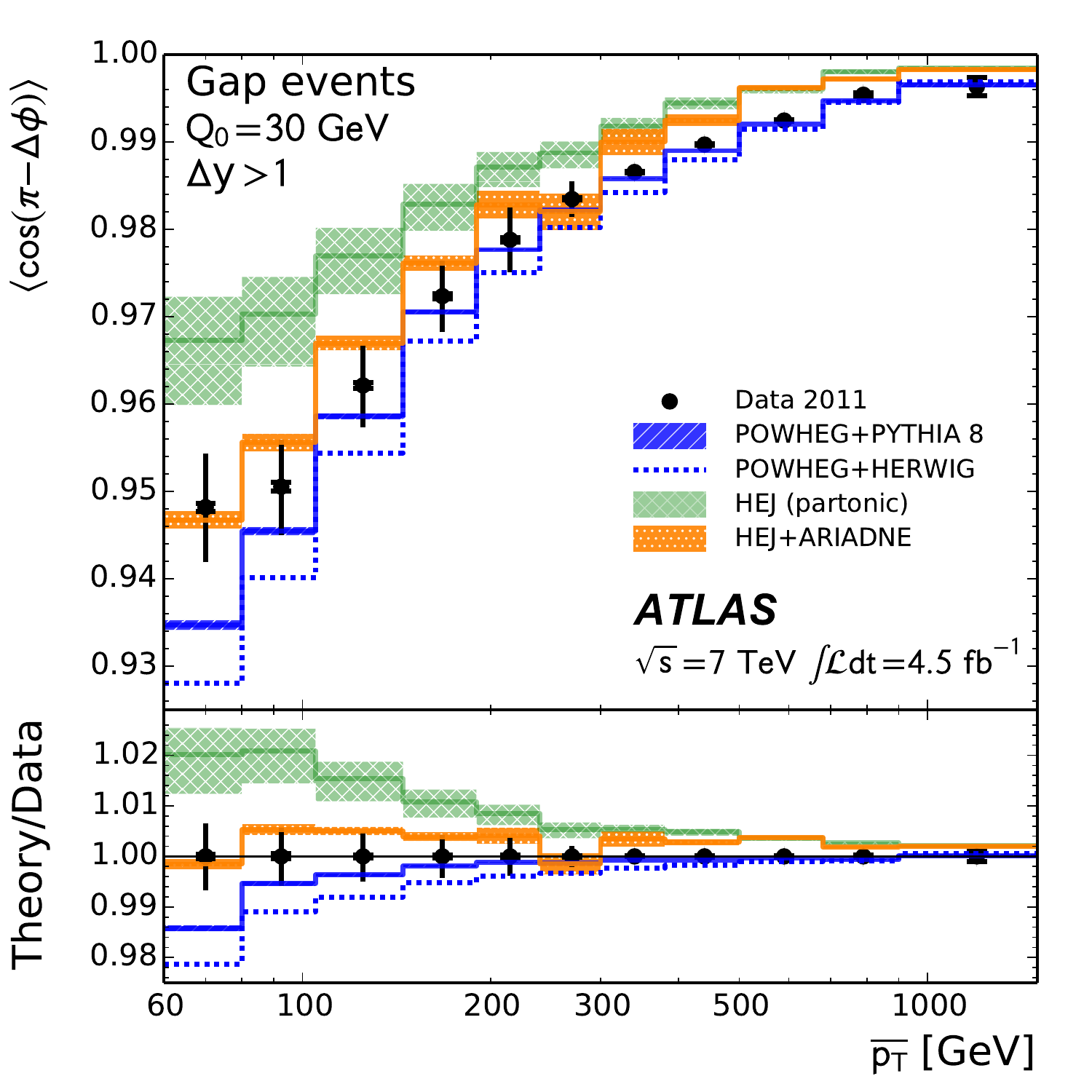}
    \label{fig:results:CosDeltaPhi_Gap_pTbarBins}}
  \\
  \subfloat[]{
    \includegraphics[width=\paperdoublefig]{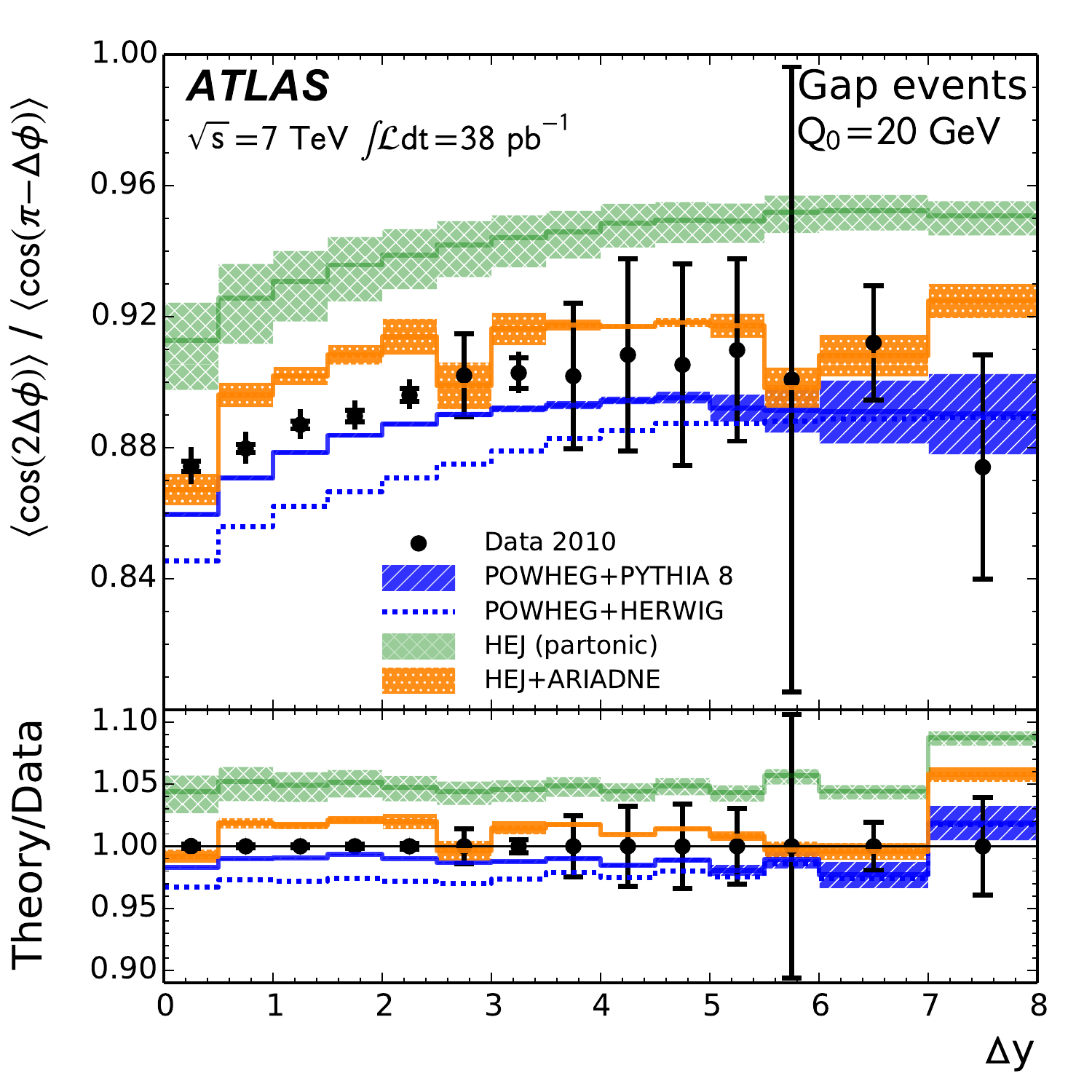}
    \label{fig:results:C2C1_Gap_dyBins}}
  \qquad
  \subfloat[]{
    \includegraphics[width=\paperdoublefig]{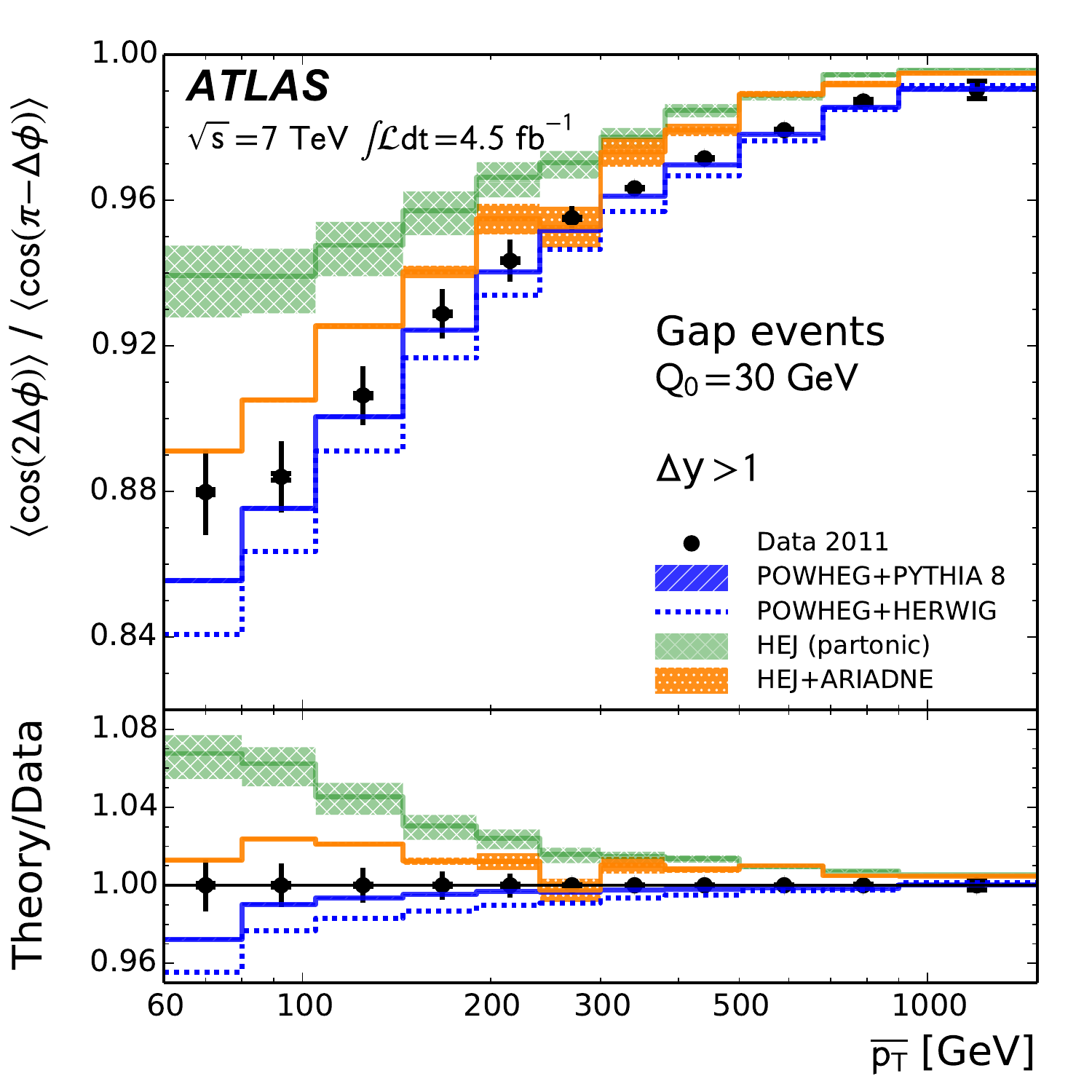}
    \label{fig:results:C2C1_Gap_pTbarBins}}
  \caption{The measured \protect\subref{fig:results:CosDeltaPhi_Gap_dyBins}\protect\subref{fig:results:CosDeltaPhi_Gap_pTbarBins} \meanCosDeltaPhi and \protect\subref{fig:results:C2C1_Gap_dyBins}\protect\subref{fig:results:C2C1_Gap_pTbarBins} \CtwoCone distributions, for gap events as a function of \protect\subref{fig:results:CosDeltaPhi_Gap_dyBins}\protect\subref{fig:results:C2C1_Gap_dyBins} \meanCosDeltaPhi \DeltaY and \protect\subref{fig:results:CosDeltaPhi_Gap_pTbarBins}\protect\subref{fig:results:C2C1_Gap_pTbarBins} \meanCosDeltaPhi \pTbar. The veto scale is \Qnought = 20 \unit{(30)}{\GeV} for data collected during 2010 (2011). For comparison, the \HEJ, \HEJA, \PowhegP and \PowhegH predictions are presented in the same way as \figureRef{fig:results:GapFraction}.}
  \label{fig:results:CosDeltaPhi_C2C1_Gap}
\end{figure*}

\FigureRef{fig:results:CosDeltaPhi_C2C1_Gap} shows the corresponding \meanCosDeltaPhi and \CtwoCone distributions for events that pass the veto requirement on additional jet activity in the rapidity interval bounded by the dijet system.
In this case, with the jet veto suppressing additional quark and gluon radiation, the spectra
show the opposite behaviour, namely a slight increase in correlation with \DeltaY, which now agrees with the rise seen in the \pTbar distribution.
This can be explained by considering that as \DeltaY or \pTbar increase, the veto requirement imposes an increasingly back-to-back topology on the \dijet system.
The spread of theoretical predictions is again large in each distribution, with the \Powheg predictions having too much decorrelation and \HEJ predicting too little decorrelation.

The use of the \Ariadne parton shower again brings \HEJ into better agreement with the data, although not as well as in the inclusive case.
The best agreement is given by \PowhegP, especially in the highest \pTbar bins.
\HEJ agrees less well with the data than in the inclusive case, showing that this region of widely separated hard jets without additional radiation in the event is not well reproduced by the \HEJ calculation.
A quantitative statement about the degree of agreement seen here between \HEJA and the data cannot, however, be made in the absence of theoretical uncertainties on this calculation.

\begin{figure*}[htpb]
  \centering
  \subfloat[]{
    \includegraphics[width=\paperdoublefig]{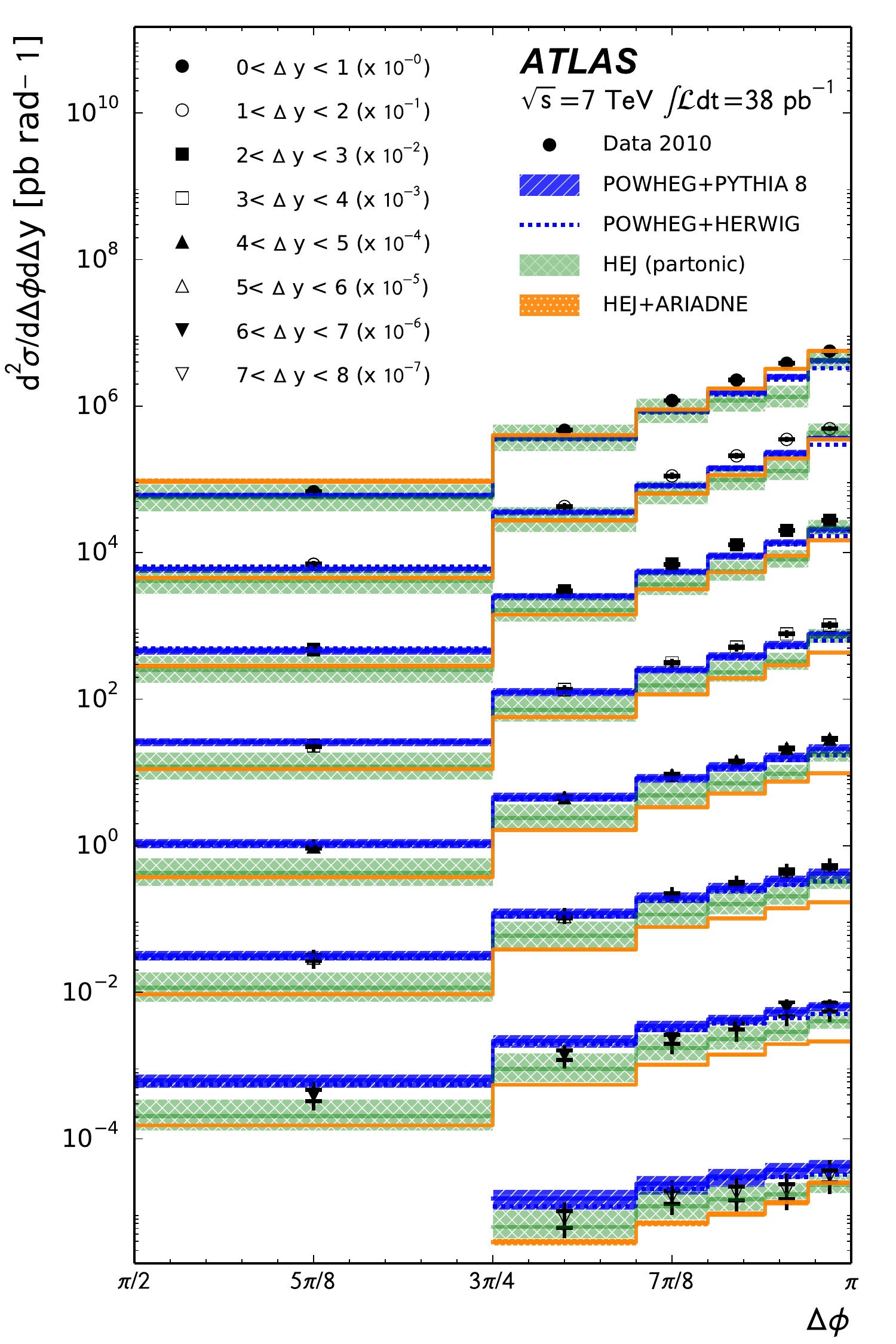}
    \label{fig:results:xs_inclusive_dySlices_dPhiBins_stack}}
  \qquad
  \subfloat[]{
    \includegraphics[width=\paperdoublefig]{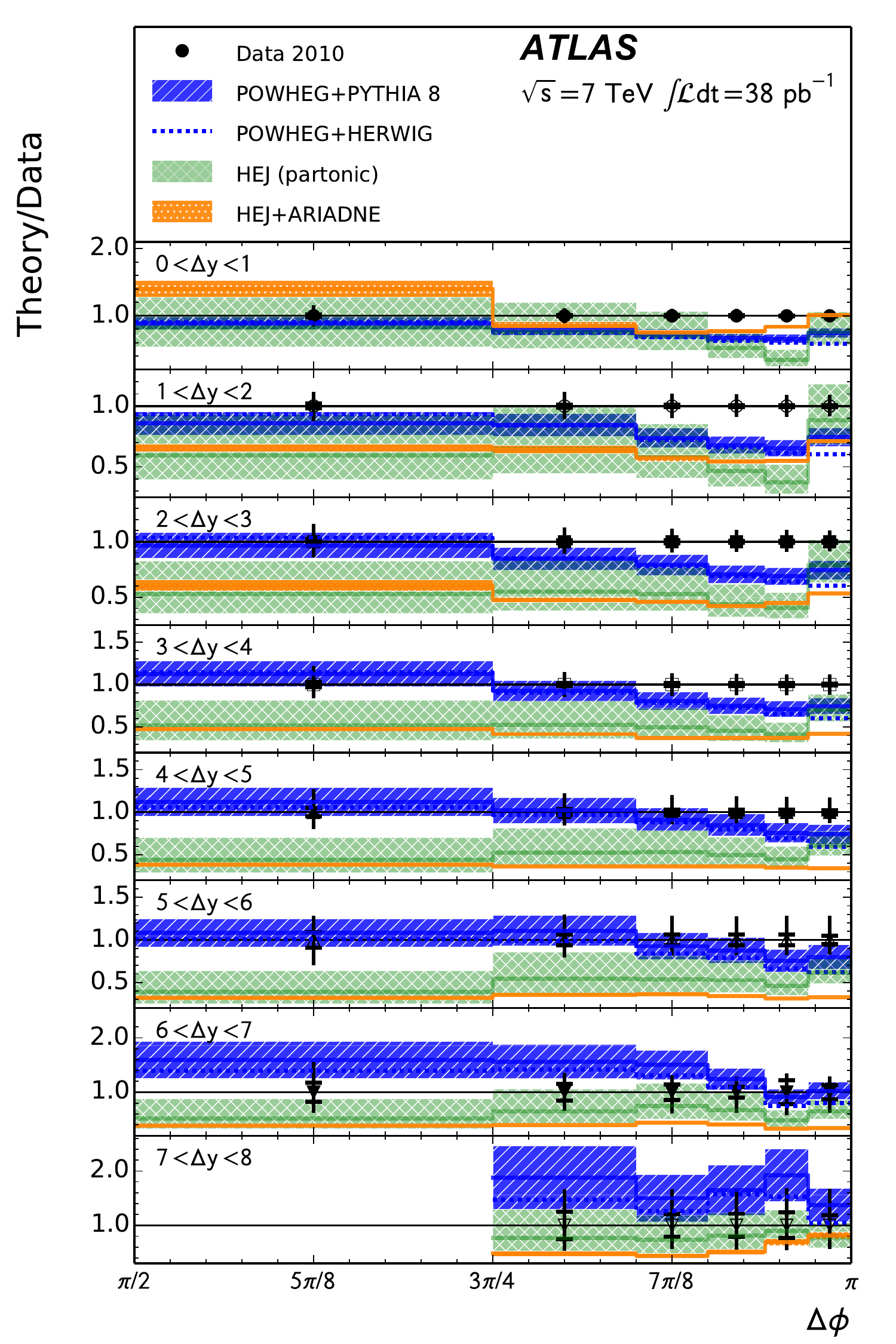}
    \label{fig:results:xs_inclusive_dySlices_dPhiBins_stackRatio}}
  \caption{The measured double-differential \xs{s} (black points) as a function of \DeltaPhi for eight slices in \DeltaY. For comparison, the \HEJ, \HEJA, \PowhegP and \PowhegH  predictions are presented in the same way as \figureRef{fig:results:GapFraction}. In \protect\subref{fig:results:xs_inclusive_dySlices_dPhiBins_stack} the absolute comparison is shown, while in \protect\subref{fig:results:xs_inclusive_dySlices_dPhiBins_stackRatio} the ratios of the predictions to the data are shown.}
  \label{fig:results:xs_inclusive_dySlices_dPhiBins}
\end{figure*}

\begin{figure*}[htpb]
  \centering
  \subfloat[]{
    \includegraphics[width=\paperdoublefig]{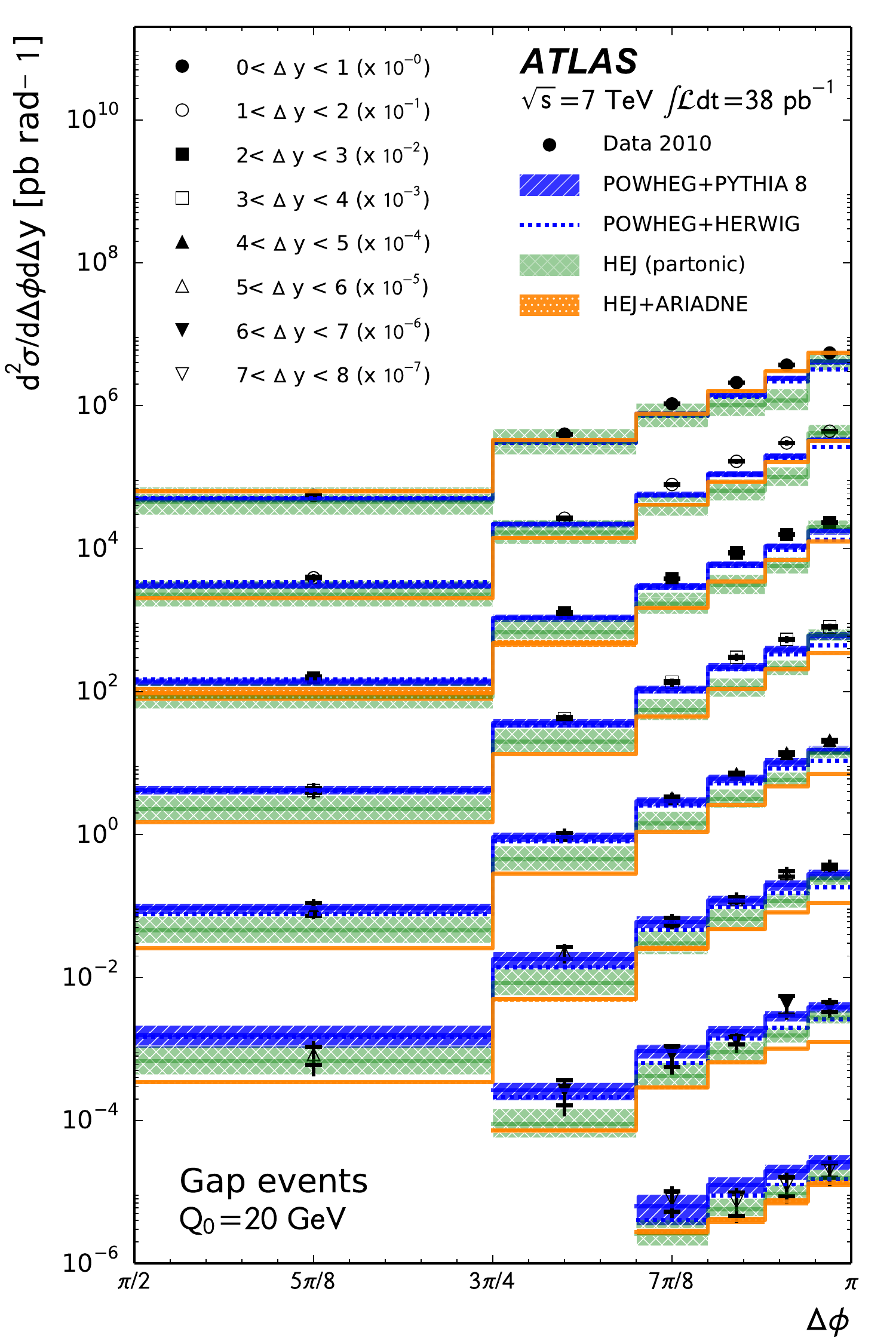}
    \label{fig:results:xs_gap_dySlices_dPhiBins_stack}}
  \qquad
  \subfloat[]{
    \includegraphics[width=\paperdoublefig]{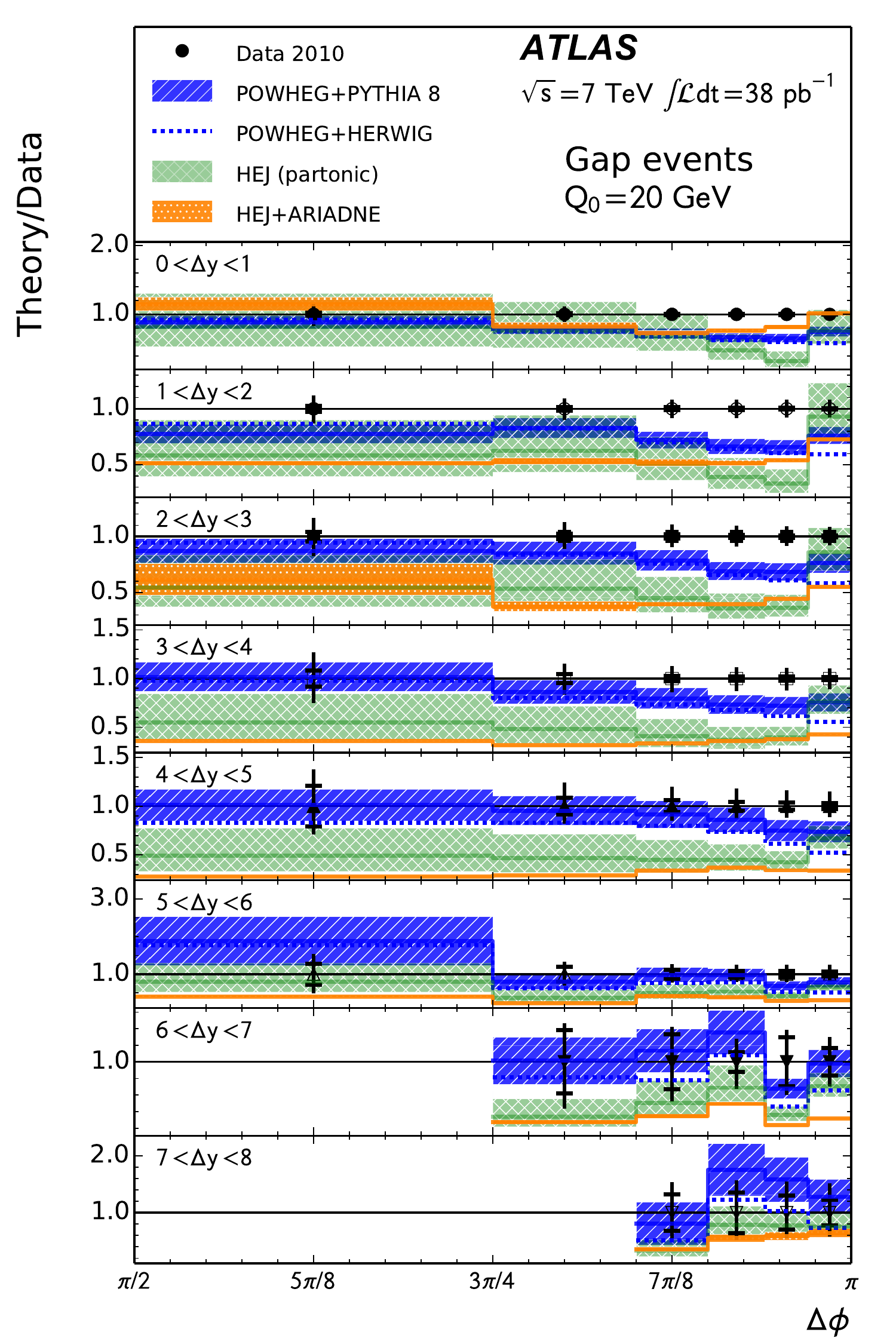}
    \label{fig:results:xs_gap_dySlices_dPhiBins_stackRatio}}
  \caption{The measured double-differential \xs{s} as a function of \DeltaPhi for eight slices in \DeltaY, for gap events. The veto scale is \maths{\Qnought = \unit{20}{\GeV}}. For comparison, the \HEJ, \HEJA, \PowhegP and \PowhegH predictions are presented in the same way as \figureRef{fig:results:GapFraction}. In \protect\subref{fig:results:xs_inclusive_dySlices_dPhiBins_stack} the absolute comparison is shown, while in \protect\subref{fig:results:xs_inclusive_dySlices_dPhiBins_stackRatio} the ratios of the predictions to the data are shown.}
  \label{fig:results:xs_gap_dySlices_dPhiBins}
\end{figure*}

\FigureRefTwo{fig:results:xs_inclusive_dySlices_dPhiBins}{fig:results:xs_gap_dySlices_dPhiBins} show the double-differential \dijet \xs{s} as functions of \DeltaPhi and \DeltaY for inclusive and gap events respectively.
The predictions from \PowhegP and \PowhegH provide a good overall description of the measured \xs{s}, within the experimental and theoretical uncertainties, with the only notable deviations occurring at high \DeltaPhi in the lowest \DeltaY bins.
This is in agreement with the observations in previous ATLAS studies~\cite{Aad:2010ad}.
\HEJ underestimates the \xs seen in data throughout the \DeltaY range, although it provides a good description of the overall shape.
This underestimate is noticeably enhanced when only gap events are considered, which is a regime far from the wide-angle, hard-emission limit for which the underlying resummation procedure is valid.

\section{Summary}\label{sec:summary}
Theoretical predictions based on perturbative QCD are tested by studying \dijet events in extreme regions of phase space.
Measurements of the gap fraction, as a function of both the rapidity separation and the average \dijet transverse momentum, together with the azimuthal decorrelation are presented as functions of \DeltaY and \pTbar, extending previous studies up to eight rapidity units in \DeltaY and \unit{1.5}{\TeV} in \pTbar.
The measurements are used to investigate the predicted breakdown of DGLAP evolution and the appearance of BFKL effects by comparing the data to the all-order resummed leading-logarithmic calculations of \HEJ and the full next-to-leading-order calculations of \Powheg.
The full data sample collected with the ATLAS detector in \unit{7}{\TeV} $pp$ collisions at the LHC is used, corresponding to integrated luminosities of \unit{\maths{36.1}}{\ipb} and \unit{\maths{4.5}}{\ifb} for data collected during 2010 and 2011 respectively.

The data show the expected behaviour of a reduction of gap events, or equivalently, an increase in jet activity, for large values of \pTbar and \DeltaY, together with an associated rise in the number of jets in the rapidity interval.
The azimuthal moments show an increase in correlation with increasing \pTbar and an increase in (de)correlation with increasing \DeltaY for gap (all) events.
The expected increase in \xs with \DeltaPhi is also seen.

The \PowhegP prediction provides a reasonable description of the data in most distributions, but shows disagreement in some areas of phase-space, particularly for the inclusive azimuthal distributions in the limit of large \DeltaY or small \pTbarQnought.
When the \Herwig parton shower is used instead of \PythiaEight, the agreement with data worsens as \Herwig predicts too many jets above the veto scale.
The partonic \HEJ prediction provides a poor description of the data in most of the distributions presented, with the exception of the gap fraction and jet multiplicity distributions as a function of \DeltaY.
The addition of the \Ariadne parton shower, which accounts for some of the soft and collinear terms ignored in the \HEJ approximation, brings the prediction closer to \PowhegP.

No single theoretical prediction is able to simultaneously describe the data over the full phase-space region considered here; in general, however, the best agreement is given by \PowhegP and \HEJA.
The variable best able to discrimate between the DGLAP-like prediction from \PowhegP and the BFKL-like prediction from \HEJ is \CtwoCone.
Here, it can clearly be seen that neither of these predictions describe the data accurately as a function of either \DeltaY or \pTbar.
It should be noted, however, that when the inclusive event sample is considered, the \HEJA model, a combination of BFKL-like parton dynamics with the colour-dipole cascade model, provides a good description of \CtwoCone in the large \DeltaY and small \pTbar regions, where the \Powheg models fail to describe the data.

In most of the phase-space regions presented, the experimental uncertainty is smaller than the spread of theoretical predictions.
These disparities between predictions represent a genuine difference in the modelling of the underlying physics and the data can, therefore, provide a crucial input for constraining parton-shower models in the future -- particularly in the case of QCD radiation between widely separated or high transverse momentum \dijet{s}.
Improved theoretical predictions are essential before any conclusions can be drawn about the presence or otherwise of BFKL effects or colour-singlet exchange in these data.

\begin{acknowledgements}
The \HEJ and \HEJA predictions used in this paper were produced by Jeppe Andersen, Jack Medley and Jennifer Smillie.

% Acknowledgements for papers with collision data
% Version 19-Feb-2014

We thank CERN for the very successful operation of the LHC, as well as the
support staff from our institutions without whom ATLAS could not be
operated efficiently.

We acknowledge the support of ANPCyT, Argentina; YerPhI, Armenia; ARC,
Australia; BMWF and FWF, Austria; ANAS, Azerbaijan; SSTC, Belarus; CNPq and FAPESP,
Brazil; NSERC, NRC and CFI, Canada; CERN; CONICYT, Chile; CAS, MOST and NSFC,
China; COLCIENCIAS, Colombia; MSMT CR, MPO CR and VSC CR, Czech Republic;
DNRF, DNSRC and Lundbeck Foundation, Denmark; EPLANET, ERC and NSRF, European Union;
IN2P3-CNRS, CEA-DSM/IRFU, France; GNSF, Georgia; BMBF, DFG, HGF, MPG and AvH
Foundation, Germany; GSRT and NSRF, Greece; ISF, MINERVA, GIF, I-CORE and Benoziyo Center,
Israel; INFN, Italy; MEXT and JSPS, Japan; CNRST, Morocco; FOM and NWO,
Netherlands; BRF and RCN, Norway; MNiSW and NCN, Poland; GRICES and FCT, Portugal; MNE/IFA, Romania; MES of Russia and ROSATOM, Russian Federation; JINR; MSTD,
Serbia; MSSR, Slovakia; ARRS and MIZ\v{S}, Slovenia; DST/NRF, South Africa;
MINECO, Spain; SRC and Wallenberg Foundation, Sweden; SER, SNSF and Cantons of
Bern and Geneva, Switzerland; NSC, Taiwan; TAEK, Turkey; STFC, the Royal
Society and Leverhulme Trust, United Kingdom; DOE and NSF, United States of
America.

The crucial computing support from all WLCG partners is acknowledged
gratefully, in particular from CERN and the ATLAS Tier-1 facilities at
TRIUMF (Canada), NDGF (Denmark, Norway, Sweden), CC-IN2P3 (France),
KIT/GridKA (Germany), INFN-CNAF (Italy), NL-T1 (Netherlands), PIC (Spain),
ASGC (Taiwan), RAL (UK) and BNL (USA) and in the Tier-2 facilities
worldwide.

\end{acknowledgements}

\bibliographystyle{atlasBibStyleWoTitle}
\bibliography{references}

\onecolumn
\clearpage
% ATLAS Collaboration author list
% Data extracted on 06-Jun-2014 for paper reference STDM-2012-17
\begin{flushleft}
{\Large The ATLAS Collaboration}

\bigskip

G.~Aad$^{\rm 84}$,
B.~Abbott$^{\rm 112}$,
J.~Abdallah$^{\rm 152}$,
S.~Abdel~Khalek$^{\rm 116}$,
O.~Abdinov$^{\rm 11}$,
R.~Aben$^{\rm 106}$,
B.~Abi$^{\rm 113}$,
M.~Abolins$^{\rm 89}$,
O.S.~AbouZeid$^{\rm 159}$,
H.~Abramowicz$^{\rm 154}$
% ...
\bigskip
\\
% ...
$^{11}$ Institute of Physics, Azerbaijan Academy of Sciences, Baku, Azerbaijan\\
% ...
$^{84}$ CPPM, Aix-Marseille Universit{\'e} and CNRS/IN2P3, Marseille, France\\
% ...
$^{89}$ Department of Physics and Astronomy, Michigan State University, East Lansing MI, United States of America\\
% ...
$^{106}$ Nikhef National Institute for Subatomic Physics and University of Amsterdam, Amsterdam, Netherlands\\
% ...
$^{112}$ Homer L. Dodge Department of Physics and Astronomy, University of Oklahoma, Norman OK, United States of America\\
$^{113}$ Department of Physics, Oklahoma State University, Stillwater OK, United States of America\\
% ...
$^{116}$ LAL, Universit{\'e} Paris-Sud and CNRS/IN2P3, Orsay, France\\
% ...
$^{152}$ Institute of Physics, Academia Sinica, Taipei, Taiwan\\
% ...
$^{154}$ Raymond and Beverly Sackler School of Physics and Astronomy, Tel Aviv University, Tel Aviv, Israel\\
% ...
$^{159}$ Department of Physics, University of Toronto, Toronto ON, Canada\\
% ...
\end{flushleft}

\end{document}